\documentclass[sigconf,authorversion]{acmart}

\usepackage{booktabs}
\usepackage{algorithm}
\usepackage{algpseudocode}
\usepackage{bbold}
\usepackage{multirow}
\usepackage{subcaption}
\usepackage{balance}

\newcommand{\sz}[1]{\textcolor{blue}{#1}}

\usepackage{enumitem}
\setlist[description,1]{leftmargin=0.8cm}

\newtheoremstyle{mydef}
{2ex}
{2ex}
{\itshape}
{}
{\scshape}
{: }
{0.5em}
{}
\theoremstyle{mydef}

\renewcommand\vec{\mathbf}

\begin{document}


\title{Recommending Related Tables}

\author{Shuo Zhang}
\affiliation{%
  \institution{University of Stavanger}
}
\email{shuo.zhang@uis.no}

\author{Krisztian Balog}
\affiliation{%
  \institution{University of Stavanger}
}
\email{krisztian.balog@uis.no}

\begin{abstract}
Tables are an extremely powerful visual and interactive tool for structuring and manipulating data, making spreadsheet programs one of the most popular computer applications.  In this paper we introduce and address the task of \emph{recommending related tables}: given an input table, identifying and returning a ranked list of relevant tables.  One of the many possible application scenarios for this task is to provide users of a spreadsheet program proactively with recommendations for related structured content on the Web.  At its core, the related table recommendation task boils down to computing the similarity between a pair of tables.  We develop a theoretically sound framework for performing table matching.  Our approach hinges on the idea of representing table elements in multiple semantic spaces, and then combining element-level similarities using a discriminative learning model.  Using a purpose-built test collection from Wikipedia tables, we demonstrate that the proposed approach delivers state-of-the-art performance.  
\end{abstract}



\maketitle

\section{Introduction}

\begin{figure}[t]
   \centering
   \includegraphics[width=0.47\textwidth]{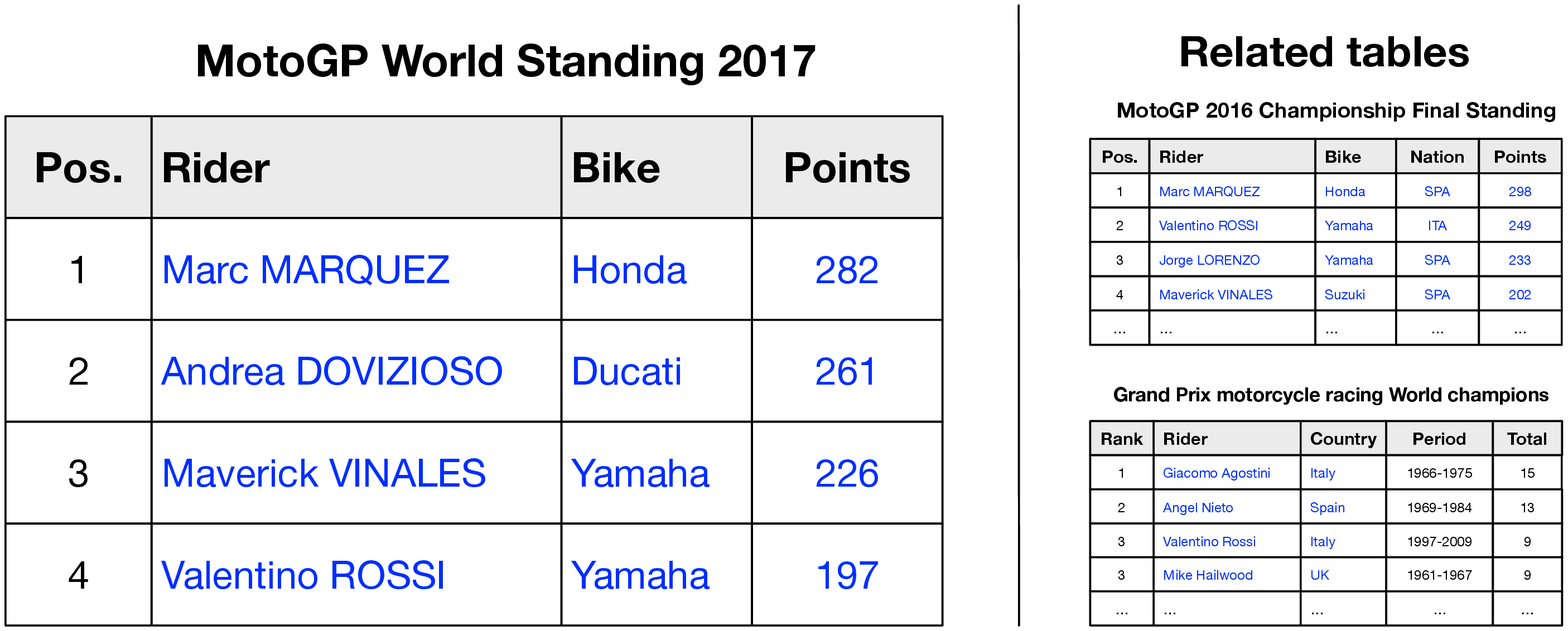} 
     \vspace*{-0.5\baselineskip}
   \caption{The task of related table recommendation is to return a ranked list of tables, given an input table.}
   \label{fig:tableretrieval}
   \vspace*{-\baselineskip}
\end{figure}

Tables are an extremely powerful visual and interactive tool for structuring and manipulating data.  The Web contains vast amounts of HTML tables~\citep{Lehmberg:2016:LPC} and there is a growing body of research utilizing relational information stored in them~\citep{Ritze:2016:PPW,Ibrahim:2016:MSE,Bhagavatula:2015:TEL,Chirigati:2016:KEU,Zhang:2018:OTG}.
Table retrieval has been recognized as an important search task ~\citep{Cafarella:2008:WEP, Zhang:2018:AHT, Pimplikar:2012:ATQ}. In this paper, we propose and address the task of \emph{recommending related tables}: given an input relational table, identify and return a ranked list of relevant tables that contain novel information (additional entities and/or attributes).  The main difference from previous work is that instead of requiring the user to express her information need explicitly, by formulating a keyword query, we can proactively recommend related tables based on any existing table as input.  This input table may be an incomplete table the user currently works on or a complete table that can be found on the Web.
Figure~\ref{fig:tableretrieval} illustrates the idea. 
Table recommendations could be helpful, for example, in equipping spreadsheet applications with a smart assistance feature for finding related content.  Alternatively, it could be implemented as a browser plugin that can be activated upon encountering a table on a webpage to find related tables (e.g., for comparison or fact validation).
%

At its core, the related table recommendation task boils down to computing a similarity score between a pair of tables, the input and candidate tables, which we shall refer to as \emph{table matching}.  
We note that table matching is a core component in many other table-related tasks beyond search and recommendation, such as table augmentation~\citep{Lehmberg:2015:MSJ,DasSarma:2012:FRT,Yakout:2012:IEA,Ahmadov:2015:THI}, question answering on tables~\citep{Pimplikar:2012:ATQ}, and table interpretation~\citep{Venetis:2011:RST, Bhagavatula:2015:TEL}.
Previous approaches may be divided into two main categories:
(i) extracting a keyword query from certain table elements and scoring candidate tables using that query, e.g., \citep{Lehmberg:2015:MSJ, Ahmadov:2015:THI}
(ii) splitting both the query and candidate tables into several elements and performing element-wise matching, e.g., \citep{DasSarma:2012:FRT, Yakout:2012:IEA, Nguyen:2015:RSS}. 
Commonly considered table elements include table caption, table entities, column headings, and table data (cell values).

Existing approaches for table matching suffer from three main shortcomings.
First, they rely on ad hoc similarity measures, tailor-made for each table element.  
Second, even though multiple table elements (caption, column headings, cell values, etc.) have been considered, a principled way of combining these element-level similarities is lacking, along with a systematic assessment of the contribution of the various table elements in such a combination.
Third, the possibility of matching elements of different types has not been explored yet (e.g., comparing the input table's headers against the candidate table's cell values).
Motivated by the above issues, our the main research objective is to develop an effective and theoretically sound table matching framework for measuring and combining table element level similarity, without resorting to hand-crafted features.
We propose an element-oriented table matching framework that hinges on the idea of representing table elements in multiple semantic spaces. 
We develop multiple methods for measuring the similarity between table elements; these are applicable to both elements of the same type (element-wise matching) and of different types (cross-element matching).  Finally, we combine element-level similarities using a discriminative learning approach. 
Through our experimental evaluation, we seek to answer the following specific research questions:
\begin{description}
	\item[RQ1] Which of the semantic representations (word-based, graph-based, or entity-based) is the most effective for modeling table elements? 
	\item[RQ2] Which of the two element-level matching strategies performs better, element-wise or cross-element?
	\item[RQ3] How much do different table elements contribute to recommendation performance? 
\end{description}


\noindent
For experimental evaluation, we develop a test collection based on Wikipedia tables.  We first present a feature-based method that combines numerous hand-crafted element-level similarity measures from the literature in a discriminative learning approach.  This method, referred to as HCF, improves over the best existing baseline method by almost 30\% in terms of NDCG@10.
We then show that our proposed novel approach, termed CRAB, based on element-level semantic representations and matching, performs on par with this strong combination of hand-crafted features.  
Our analysis reveals that cross-element matching, while seemingly unintuitive, can indeed be beneficial.  We further demonstrate that recommendation performance increases as the input table grows, either horizontally or vertically, which attests to the capability of our table matching framework to effectively utilize larger inputs.
In summary, this paper makes the following contributions:

\begin{itemize}
	\item We introduce and address the related table recommendation task, adapt existing methods, and present a discriminative approach that combines hand-crafted features from the literature (Sect.~\ref{sec:baseline}). 
	\item We develop a general a table matching framework and specific instantiations of this framework (Sect.~\ref{sec:our}).
	\item We construct a purpose-built test collection (Sect.~\ref{sec:tc}), perform a thorough experimental evaluation, and provide valuable insights and analysis (Sect.~\ref{sec:eval}).
\end{itemize}
The resources developed within this paper will be made publicly available upon acceptance.

\vspace*{-0.5\baselineskip}
\section{Using Hand-Crafted Features}
\label{sec:baseline}

We present an approach, termed HCF, which combines various table similarity measures from the literature in a feature-based ranking framework.  Additionally, we introduce a set of features to describe the input and candidate tables on their own. 
As we will show in our experimental section, this approach outperforms the best method from the literature by almost 30\%.
Therefore, even though the individual features are not regarded as novel, the rich feature set we introduce here does represent an important contribution.  
 
\vspace*{-0.5\baselineskip}
\subsection{Recommender Framework}
\label{sec:baseline:rf}

The objective of table matching is to compute the similarity between an input table $\tilde{T}$ and a candidate table $T$, expressed as $\mathit{score}(\tilde{T},T)$. 
Formally, our goal is to learn a recommender model $h(\tilde{T},T)=h(\vec{x}_{\tilde{T},T})$ that gives a real-valued score for an input and candidate table pair, or equivalently, to the corresponding feature vector $\vec{x}_{\tilde{T},T}$.  The feature vector is defined as:
\begin{align}
	\vec{x}_{\tilde{T},T} = & \big\langle \phi_1(\tilde{T}), \dots, \phi_n(\tilde{T}), 	\label{eq:feat} \\
	& \phi_{n+1}(T), \dots, \phi_{2n}(T), \nonumber \\
	& \phi_{2n+1}(\tilde{T}, T), \dots, \phi_{2n+m}(\tilde{T}, T) \big\rangle  \nonumber 
\end{align}
There are two main groups of features.
The first $2n$ features are based on the characteristics of the input and candidate tables, respectively ($n$ features each). These features are discussed in Sect.~\ref{sec:baseline:tablefeat}.
Then, $m$ features are used for representing the similarity between a pair of tables; these are described in Sect.~\ref{sec:baseline:tablesim}.

\vspace*{-0.5\baselineskip}
\subsection{Table Similarity Features}
\label{sec:baseline:tablesim}

On the high level, all the existing methods  operate by (i) subdividing tables into a number of \emph{table elements}, such as page title ($T_p$), table caption ($T_c$), table topic ($T_t$), column headings ($T_H$), table entities ($T_E$), and table data ($T_D$), (ii) measuring the similarity between various elements of the input and candidate tables, and (iii) combining these element-level similarities into a final score. 
We adapt all element-level similarity scores from the individual methods (detailed in Sect.~\ref{sub:er}) as table similarity features.  These are shown in Table~\ref{tbl:features_tsf}, grouped by table elements.


\begin{table}[t]
\centering
\small
\caption{Table similarity features. All values are in $[0,1]$. }
\vspace*{-0.75\baselineskip}
\begin{tabular}{p{0.1cm}p{6.5cm}c}
	\toprule
	\multicolumn{2}{l}{\textbf{Element / Feature}} & \textbf{Source} \\ 
	\midrule
	\multicolumn{3}{l}{\emph{Page title} ($\tilde{T}_p \leftrightarrow T_p$)} \\ 
	& InfoGather page title IDF similarity score & \cite{Yakout:2012:IEA} \\
	\midrule
	\multicolumn{3}{l}{\emph{Table headings} ($\tilde{T}_H \leftrightarrow T_H$)} \\ 
	& MSJE heading matching score  & \cite{Lehmberg:2015:MSJ} \\
	& Schema complement schema benefit score   & \cite{DasSarma:2012:FRT} \\
	& InfoGather heading-to-heading similarity & \cite{Yakout:2012:IEA} \\
	& Nguyen et al. heading similarity  & \cite{Nguyen:2015:RSS} \\
	\midrule
	\multicolumn{3}{l}{\emph{Table data} ($\tilde{T}_D \leftrightarrow T_D$)} \\ 
	& InfoGather column-to-column similarity & \cite{Yakout:2012:IEA} \\
	& InfoGather table-to-table similarity & \cite{Yakout:2012:IEA} \\
	& Nguyen et al. table data similarity  & \cite{Nguyen:2015:RSS} \\
	\midrule
	\multicolumn{3}{l}{\emph{Table entities} ($\tilde{T}_E \leftrightarrow T_E$)} \\ 
	 & Entity complement entity relatedness score  & \cite{DasSarma:2012:FRT}  \\ 
	 & Schema complement entity overlap score  &  \cite{DasSarma:2012:FRT} \\
	\bottomrule
\end{tabular}
\label{tbl:features_tsf}
\vspace*{-1\baselineskip}
\end{table}

\vspace*{-0.5\baselineskip}
\subsection{Table Features}
\label{sec:baseline:tablefeat}

Additionally, we present a set of features that characterize individual tables.  
Table features are computed for both the input and candidate tables.  They might be thought of as analogous to the query and document features, respectively, in document retrieval~\citep{Macdonald:2012:UQF}.
In fact, we adapt some features from document retrieval, such as query IDF score~\citep{Qin:2010:LBC}. 
Specifically, we compute IDF for the table caption and page title elements, by summing up the term IDF scores: $IDF(f)=\sum_{t \in f}IDF(t)$.
We further consider general table descriptors from ~\cite{Bhagavatula:2013:MEM}, like the number of table rows, columns, and empty cells.  Another group of features is concerned with the page in which the table is embedded. The includes page connectivity (inLinks and outLinks), page popularity (page counts), and the table's importance within the page (tableImportance and tablePageFraction).  Table~\ref{tbl:features_table} provides an overview of table features.

\vspace*{-0.5\baselineskip}
\section{The CRAB Approach}
\label{sec:our}

This section presents our novel approach for table matching.  Our contributions are twofold. We introduce a general element-oriented table matching framework in Sect.~\ref{sec:our:fw} followed by specific instantiations of this framework, referred to as CRAB, in Sect.~\ref{sec:our:crab}.

\vspace*{-0.5\baselineskip}
\subsection{Element-Level Table Matching Framework}
\label{sec:our:fw}

We combine multiple table quality indicators and table similarity measures in a discriminative learning framework.  Input and candidate table pairs are described as a feature vector, shown in Eq.~\eqref{eq:feat}.
The main novelty lies in how table similarity is estimated.  Instead of relying on hand-crafted features, like the ones presented in Sect.~\ref{sec:baseline}, we represent table elements in a uniform manner.  Moreover, instead of relying of lexical matches, we perform the matching of table elements in multiple semantic spaces.

Let $\tilde{T}_{x1}^y$ denote element $x1$ of the input table $\tilde{T}$ in representation space $y$.  Similarly, let $T_{x2}^y$ denote element $x2$ of the candidate table $T$ in representation space $y$.  We then take table similarity features to be element-level matching scores:
\begin{equation*}
	\phi_i(\tilde{T}, T) = \mathit{sim}(\tilde{T}_{x1}^y, T_{x2}^y) ~,
\end{equation*}
where $i \in [2n+1,2n+m]$ and $\mathit{sim}()$ is a similarity function. Importantly, these similarity functions are applicable both to elements of the same type ($x1=x2$), referred to as \emph{element-wise matching} (e.g., caption vs. caption, headings vs. headings, etc.) and to elements of different types ($x1 \neq x2$), referred to as \emph{cross-element matching} (e.g., caption vs. headings or headings vs. data).
Next, we present various ways of representing table elements (Sect.~\ref{sec:our:fw:repr}), and measuring element-level similarity (Sect.~\ref{sec:our:fw:sim}).  

\vspace*{-0.25\baselineskip}
\subsubsection{Representing Table Elements}
\label{sec:our:fw:repr}

Each table element, $T_x$, is represented both in a \emph{term space} and in a \emph{semantic space}.
We start with the former one. $T_x$ is described as a weighted vector of terms, where terms may be words or entities. Formally, $T_{\vec{x}} = [ t_1,\dots,t_N ]$, where $t_i$ corresponds to the weight of the $i$th term in the vocabulary. For words, the vocabulary is the set of unique words in the table corpus, for entities it is the set of entries in a knowledge base. 
We also represent each table element in a semantic space.  
Given a semantic space $y$, each term $t_i$ is described by a corresponding \emph{embedding vector}, $\vec{t}_i^y$.  The space of embeddings may be words, entities, or graphs (cf. Sect.~\ref{sec:our:crab:elements}).

\begin{figure}[tbp]
	\centering
	\includegraphics[width=0.30\textwidth]{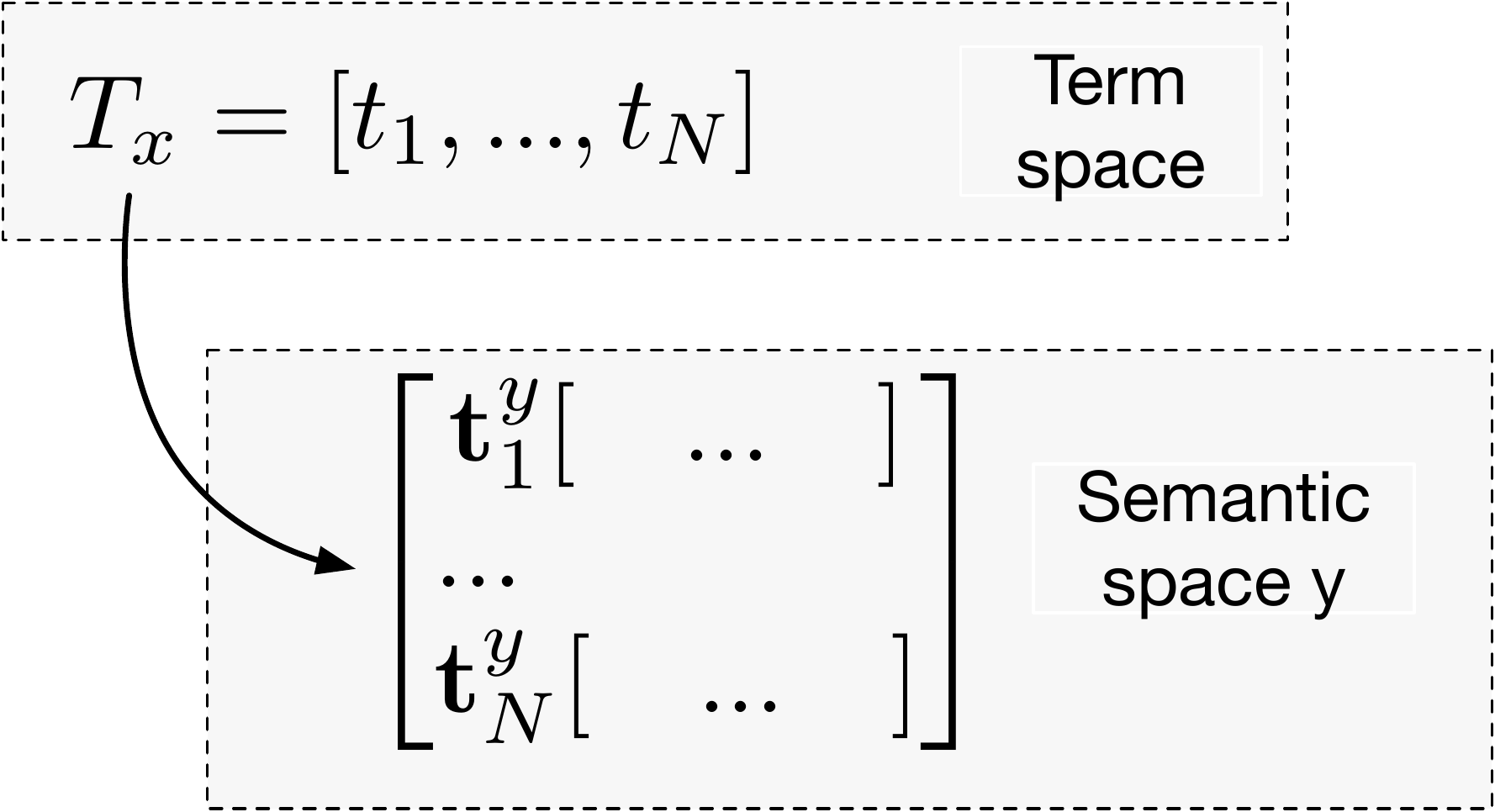}
	\caption{Representation of a table element $T_x$ in the term and in a given semantic space $y$.}
	 \label{fig:repr}
\end{figure}

In summary, each table element is represented in the term space by a term vector $T_{\vec{x}}$, and each term $t_i \in T_{\vec{x}}$ is represented by a semantic vector $\vec{t}_i^y$.
Note that the term space serves only as an intermediate representation, to help map table elements to semantic space $y$. The subsequent element-level matching will only be performed in this semantic space.
See Fig.~\ref{fig:repr} for an illustration.

\begin{table}[t]
\centering
\small
\caption{Table features.}
\vspace*{-0.75\baselineskip}
\begin{tabular}{p{2cm}p{4.5cm}l}
	\toprule
	\textbf{Feature} & \textbf{Description} & \textbf{Source} \\
	\midrule
	\#rows & Number of rows in the table & \cite{Cafarella:2008:WEP,Bhagavatula:2013:MEM} \\
	\#cols & Number of columns in the table & \cite{Cafarella:2008:WEP,Bhagavatula:2013:MEM}  \\
	\#of NULLs & Number of empty table cells & \cite{Cafarella:2008:WEP,Bhagavatula:2013:MEM} \\
	IDF($T_c$) & Table caption IDF & \cite{Qin:2010:LBC} \\	
	IDF($T_p$) & Table page title IDF & \cite{Qin:2010:LBC} \\	
	inLinks & Number of in-links to the page embedding the table & \cite{Bhagavatula:2013:MEM} \\
	outLinks & Number of out-links from the page embedding the table & \cite{Bhagavatula:2013:MEM} \\
	pageViews & Number of page views & \cite{Bhagavatula:2013:MEM} \\
	tableImportance & Inverse of number of tables on the page & \cite{Bhagavatula:2013:MEM} \\
	tablePageFraction & Ratio of table size to page size & \cite{Bhagavatula:2013:MEM} \\
	\bottomrule
\end{tabular}
\vspace*{-0.5\baselineskip}
\label{tbl:features_table}
\end{table}

\vspace*{-0.25\baselineskip}
\subsubsection{Measuring Element-level Similarity}
\label{sec:our:fw:sim}

\begin{figure}[tbp]
    \begin{subfigure}[b]{0.225\textwidth}
        \centering
        \includegraphics[width = 1\textwidth]{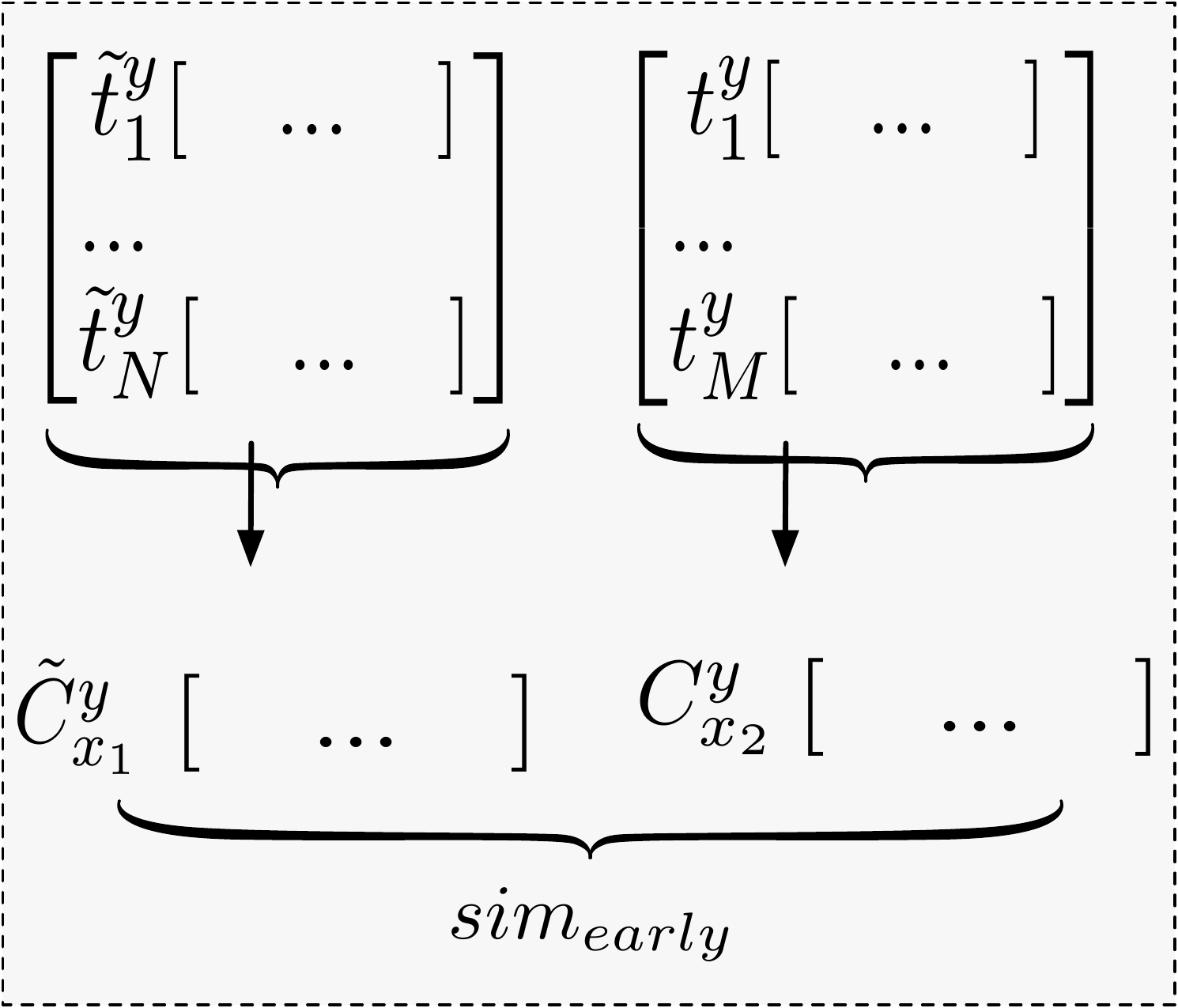}
    \end{subfigure}
    \begin{subfigure}[b]{0.225\textwidth}
        \centering
        \includegraphics[width = 1\textwidth]{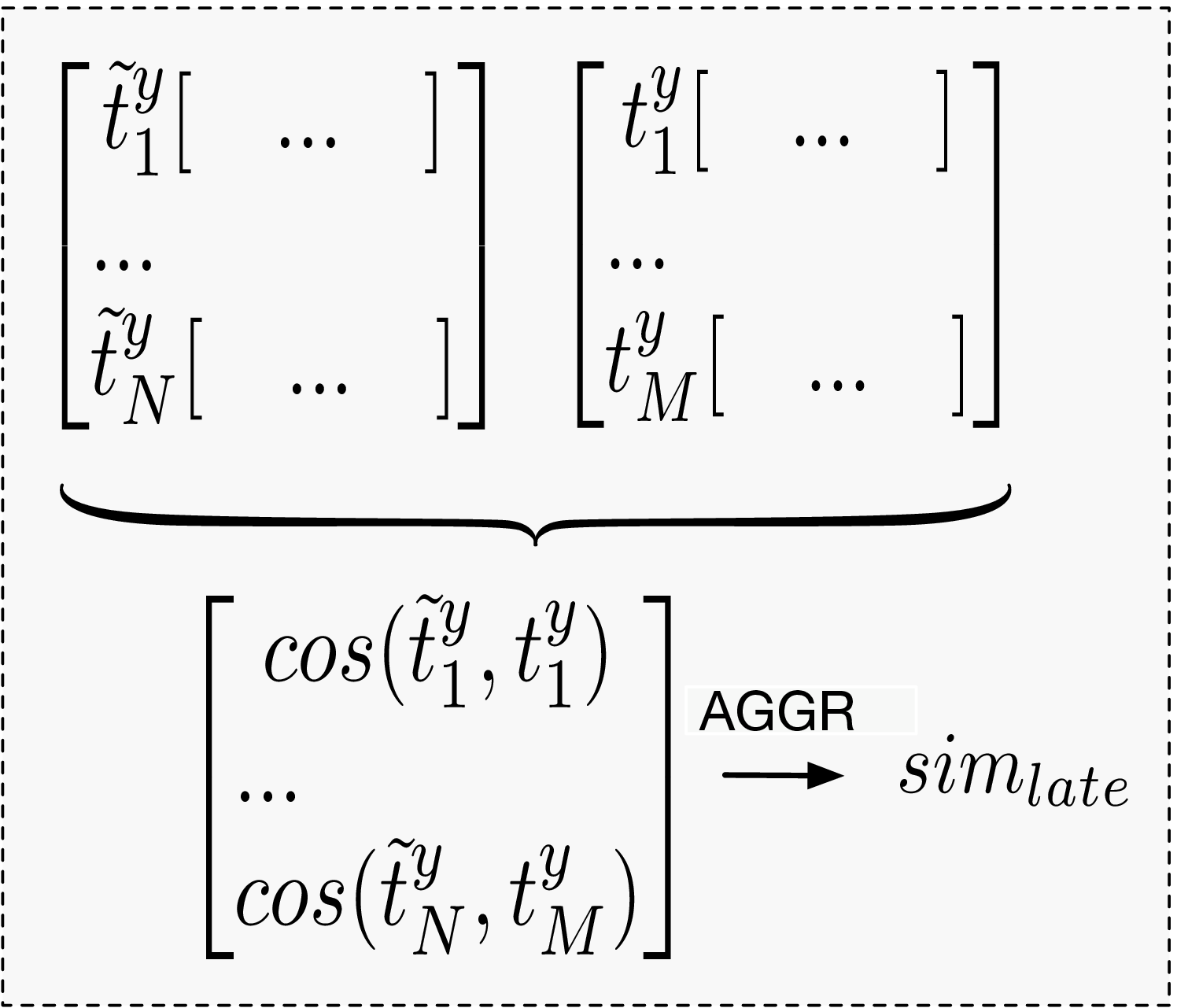}       
    \end{subfigure}
    \caption{Illustration of element-level similarity methods.}
\label{fig:early_late}
\end{figure}

We estimate the similarity between two table elements, $\tilde{T}_{x_1}$ and $T_{x_2}$, based on their semantic representations. 
Notice that these semantic representations (that is, the embedding vectors $\vec{t}_i^y$) are on the term level and not on the element level.  Thus, the term embedding vectors need to be aggregated on the element level.  Inspired by our previous work~\citep{Zhang:2017:ATR}, we present four specific element-level similarity methods.  These are roughly analogous to the early and late fusion strategies in~\cite{Snoek:2005:EVL,Zhang:2017:DPF}. We refer to Fig.~\ref{fig:early_late} for an illustration.

One strategy, referred to as \emph{early fusion}, represents each table element $T_x$ in semantic space $y$ by combining the term-level semantic vectors to a single element-level semantic vector, $\vec{C}_x^y$.  We take the weighted centroid of term-level semantic vectors:


%
\begin{equation*}
	\vec{C}_x^y[i] = \frac{\sum_{j=1}^N t_j \times \vec{t}_j^y[i] }{\sum_{j=1}^N t_j} ~,
\end{equation*}
where $[i]$ refers to the $i$th element of the vector.
Then, the similarity of two table elements is taken to be the cosine similarity of their respective centroid vectors:
\begin{equation*}
	sim_{early}(\tilde{T}_{x_1}, T_{x_2}) = \cos(\vec{C}_{x1}^y, \vec{C}_{x2}^y) ~.
\end{equation*}
According to another strategy, referred to as \emph{late fusion}, we first compute the cosine similarities between all pairs of semantic vectors.  Then, these term-level similarity scores are aggregated into an element-level score:
\begin{equation*}
	sim_{late}(\tilde{T}_{x_1}, T_{x_2}) = \mathit{aggr}(\{\cos(\vec{t}_1,\vec{t}_2): \vec{t}_1 \in \tilde{T}_{\vec{x_1}}^{y}, \vec{t}_2 \in T_{\vec{x_2}}^{y} \}) ~,
\end{equation*}
where $\mathit{aggr}()$ is an aggregation function.
Specifically, we use $\mathit{max}()$, $\mathit{sum}()$, and $\mathit{avg}()$ as aggregation functions.

\vspace*{-0.5\baselineskip}
\subsection{CRAB}
\label{sec:our:crab}

We detail a specific instantiation of our framework, which includes the representation of table elements (Sect.~\ref{sec:our:crab:elements}) and the element-level similarity scores that are used as ranking features (Sect.~\ref{sec:our:crab:features}).

%

\vspace*{-0.25\baselineskip}
\subsubsection{Representing Table Elements}
\label{sec:our:crab:elements}

We split tables into the following elements and represent them in (at most) two term spaces, words and entities, as follows:
\begin{itemize}
    \item \textbf{Table headings} ($T_H$) Table headings are represented only as words, since entity occurrences in headings are extremely rare.  If case entities appear in headings, we assign them to the table data element.
    \item \textbf{Table data} ($T_D$) The contents of table cells are used both as words and as entities. For the latter, entity mentions need to be recognized and disambiguated; such annotations be made readily available as markup (e.g., in Wikipe\-dia tables) or may be obtained automatically using entity linking techniques~\citep{Shen:2015:ELK}.
    \item \textbf{Table topic} ($T_t$) For simplicity, we combine table caption and page title into a single \emph{table topic} element.  We can directly use this text for representing the table topic in the word space.  To obtain entities for the table topic, we use the table topic text as a search query to retrieve the top-$k$ relevant entities from a knowledge base. Specifically, we employ the MLM~\cite{Hasibi:2017:NTE} retrieval method with $k=10$.
    \item \textbf{Table entities} ($T_E$) Many relational tables have a core entity column~\cite{Bhagavatula:2015:TEL,Venetis:2011:RST}, while the rest of columns represent attributes of those entities.  For this table element we only keep entities from the table's \emph{core column}, i.e., the column with the highest entity rate.  We estimate entity rate by calculating the number of column cells that contain an entity and divide it by the number of rows. 
\end{itemize}


\noindent
We consider three semantic spaces for representing table elements: word, entity, and graph embeddings.
These are explained below. 

\begin{description}
	\item[Word embeddings] Each table element is represented in the term space as a TF-IDF-weighted vector of words.  I.e., $t_i \in T_x$ refers to the TF-IDF weight of the $i$th word in the vocabulary.  Then, each word is represented in the semantic space $y=w$ by a word embedding vector $\mathbf{t}_i^w$. Specifically, we use pre-trained Word2vec~\cite{Mikolov:2013:DRW, Pennington:2014:GGV} vectors using Google News.

	\item[Graph embeddings] Each table element is represented in the term space as a binary vector of entities.  I.e., $t_i \in T_x$ is $1$ if the $i$th entity in the knowledge base appears in table element $T_x$, and is $0$ otherwise. Then, each entity is represented in the semantic space $y=g$ by a graph embedding vector $\mathbf{t}_i^g$. Specifically, we use pre-trained Graph2vec\cite{Ristoski:2016:RGE} vectors.
	
	\item[Entity embeddings] We use the same term space representation as for graph embedding, i.e., each table element is described as a binary vector of entities. Then, each entity $t_i$ is represented in the semantic space $y=e$ as a vector of linked entities.  I.e., the dimensionality of $\mathbf{t}_i^e$ is the total number of entities in the knowledge base. The $j$th element of the related entity vector is expressed as $\mathbf{t}_i^g[j]=\mathbb{1}(e_j)$, where $\mathbb{1}$ is a binary indicator function that returns $1$ if $e_i$ and $e_j$ link to each other, otherwise returns 0.
\end{description}

\vspace*{-0.25\baselineskip}
\subsubsection{Table Similarity Features}
\label{sec:our:crab:features}

Existing methods have only considered matching between elements of the same type, referred to as \emph{element-wise} matching.  Our framework also enables us to measure the similarities between elements of different types in a principled way, referred to as \emph{cross-element} matching. Finally, as before, we can also utilize table features that characterize the input and candidate tables.
Below, we detail the set of features used for measuring element-level similarity.


\begin{description}
	\item [Element-wise similarity] We compute the similarity between elements of the same type from the input and candidate tables.  Each table element may be represented in up to three semantic spaces. Then, in each of those spaces, similarity is measured using the four element-level similarity measures (early, late-max, late-sum, and late-avg).  Element-wise features are summarized in the left half of Table~\ref{tbl:element-wise-feature}.


	\item [Cross-element similarity] This approach compares table elements of different types in an asymmetrical way. Each pair of elements need to be represented in the same semantic space. Then, the same element-level similarity measures may be applied, as before.  We list the cross-element similarity features in the right half of Table~\ref{tbl:element-wise-feature}.

\end{description}

\begin{table}[!tbp]
  \centering
  \caption{Element-wise and cross-element features used in CRAB. The dimension is $r \times \# s \times \# m$, where $r$ is reflection (1 for element-wise and 2 for cross-element), $s$ is the number of semantic spaces, and $m$ is the number of element-wise similarity measures.}
  \begin{tabular}{ l  l |l l}
    \toprule 
	\textbf{Element} & \textbf{Dimension} &  \textbf{Element} & \textbf{Dimension} \\
	\midrule
	$\tilde{T}_H$ to $T_H$          &  $1 \times 1 \times 4 = 4$ 
	& $\tilde{T}_H$ to $T_{t}$          &  $2 \times 1 \times 4 = 8$\\
	$\tilde{T}_D$ to $T_D$           &  $1 \times 3 \times 4 = 12$
	& $\tilde{T}_H$ to $T_{D}$          &  $2 \times 1 \times 4 = 8$\\
	$\tilde{T}_{E}$ to $T_{E}$      &  $1 \times 2 \times 4 = 8$ 
	&$\tilde{T}_D$ to $T_{t}$           &  $2 \times 3 \times 4 = 24$\\
	$\tilde{T}_{t}$ to $T_{t}$        &  $1 \times 3 \times 4 = 12$ 
	&$\tilde{T}_D$ to $T_{E}$           &  $2 \times 2 \times 4 = 16$\\
	& 
	& 	$\tilde{T}_{t}$ to $T_{E}$           &  $2 \times 2 \times 4 = 16$ \\
	\midrule
	Total & 36 & & 72 \\
    \bottomrule
  \end{tabular}
  \label{tbl:element-wise-feature}
\end{table}

\noindent
We present four specific instantiations of our table matching framework, by considering various combinations of the three main groups of features.  These instantiations are labelled as CRAB-1 .. CRAB-4 and are summarized in Table~\ref{tbl:crab}.

\vspace*{-0.5\baselineskip}
\section{Test collection}
\label{sec:tc}
We introduce our test collection, which consists of a table corpus, a set of query tables, and corresponding relevance assessments.

\vspace*{-0.5\baselineskip}
\subsection{Table Corpus}
\label{sec:expdesign:data}

We use the WikiTables corpus~\cite{Bhagavatula:2015:TEL}, which contains 1.6M tables extracted from Wikipedia.  
The knowledge base we use is DBpedia (version 2015-10). 
We restrict ourselves to entities which have a short textual summary (abstract) in the knowledge base (4.6M in total). 
Tables are preprocessed as follows.  For each cell that contains a hyperlink we check if it points to an entity that is present in DBpedia.  If yes, we use the DBpedia identifier of the linked entity as the cell's content (with redirects resolved); otherwise, we replace the link with the anchor text (i.e., treat it as a string).

\vspace*{-0.5\baselineskip}
\subsection{Test Tables and Relevance Assessments}
Due to the lack of standard test collections, we sample 50 Wikipedia tables from the table corpus to be used as test input cases. These tables cover a diverse set of topics, including sports, music, films, food, celebrities, geography, and politics.  Each table is required to have at least five rows and three columns~\citep{Zhang:2017:ESA}.  

Ground truth relevance labels are obtained as follows.  For each input table, three keyword queries are constructed: (i) caption, (ii) table entities (entities from table plus the entity corresponding to the Wikipedia page in which the table is embedded), and (iii) table headings.  Each keyword query is used to retrieve the top 150 results, resulting in at most 450 candidate tables for each query table.  
All methods that are compared in the experimental section operate by reranking these candidate sets.
For each method, the top 10 results are manually annotated.

Each table pair (i.e., input and recommended tables) is judged on a three point scale: non-relevant (0), relevant (1), and highly relevant (2).  A table is highly relevant if it is about the same topic as the input table, but contains additional novel content that is not present in the input table.
A table is relevant if it is on-topic, but it contains limited novel content; i.e., the content largely overlaps with that of the input table.
Otherwise, the table is not relevant; this also includes tables without substantial content.
Three colleagues were employed and trained as annotators.  We take the majority vote as the relevance label; if no majority vote is achieved, the mean score is used as the final label.  To measure inter-annotator agreement, we compute the Fleiss Kappa test statistics, which is 0.6703. According to \cite{Fleiss:1971:MNS}, this is considered as  substantial agreement.

\begin{table}[t]
\caption{Features used in various instantiations of our element-wise table matching framework. Element-wise and cross-element features are summarized in Table~\ref{tbl:element-wise-feature}, table feature are listed in Table~\ref{tbl:features_table}.}
\label{tbl:crab}
\begin{tabular}{cccc}
    \toprule  
	\multirow{2}{*}{\textbf{Method}}& \multicolumn{3}{c}{\textbf{Table similarity features}} \\
	\cline{2-4}
	                     & \textbf{Element-wise} & \textbf{Cross-element}  & \textbf{Table features}  \\
	\midrule
     CRAB-1 & $\surd$ &   & \\
     CRAB-2 & $\surd$ &   & $\surd$ \\
     CRAB-3 & & $\surd$ & $\surd$ \\ 
     CRAB-4 & $\surd$ & $\surd$ & $\surd$ \\
    \bottomrule
  \end{tabular}
 
\end{table}

\vspace*{-0.5\baselineskip}
\subsection{Evaluation Metrics}
We evaluate table recommendation performance in terms of Normalized Discounted Cumulative Gain (NDCG) at cut-offs 5 and 10.  To test significance, we use a two-tailed paired t-test and write $\dag$/$\ddag$ to denote significance at the 0.05 and 0.01 levels, respectively.

\vspace*{-0.5\baselineskip}
\section{Evaluation}
\label{sec:eval}
In this section, we report on the experiments we conducted to answer our research questions.

\subsection{Baselines}
\label{sub:er}

We implement eight existing methods from literature as baselines. The table elements used in these methods are listed in Table~\ref{tbl:baseline_method}.

\begin{description}
	\item[Keyword-based search using $T_E$] 
 The candidate table's score is computed by taking the terms from $\tilde{T}_E$ as the keyword query~\citep{Ahmadov:2015:THI}.
	\item[Keyword-based search using $T_H$]	\citet{Ahmadov:2015:THI} also use table headings as keyword queries.
	\item[Keyword-based search using $T_c$]	 Additionally, in this paper we also consider using the table caption as a query.
	\item[Mannheim Search Join Engine]	All candidate tables are scored against the input table using the FastJoin matcher~\citep{Wang:2011:FEM}.
	\item[Schema complement]	 \citet{DasSarma:2012:FRT} aggregate the benefits of adding additional attributes from candidates tables to input tables as the matching score.
	\item[Entity complement]	 The aggregated scores of the benefits of adding additional entities is taken as the matching score~\citep{DasSarma:2012:FRT} .
	\item[Nguyen et al.]	 Headings and table data are represented as term vectors for table matching in~\citep{Nguyen:2015:RSS}.
	\item[InfoGather] Element-wise similarity across four table elements: table data, column values, page title, and column headings are combined by training a linear regression scorer~\citep{Yakout:2012:IEA}.		
\end{description}

\sz{
\begin{table}[t]
  \centering
  \caption{Table elements used in existing methods.}
  \begin{tabular}{lc@{~~~~~}c@{~~~~~}c@{~~~~~}c@{~~~~~}c}
    \toprule
    \textbf{Method}                                                 & \textbf{$T_c$}    & \textbf{$T_p$}  & \textbf{$T_E$} & \textbf{$T_H$} & \textbf{$T_D$} \\
    \midrule
    Keyword-based search using $T_E$~\citep{Ahmadov:2015:THI} &         &         & $\surd$ &         &         \\
    Keyword-based search using $T_H$~\citep{Ahmadov:2015:THI} &         &         &         & $\surd$ &         \\
    Keyword-based search using $T_c$ & $\surd$ &         &         &         &         \\
    Mannheim Search Join Engine~\citep{Wang:2011:FEM}        &         &         &         & $\surd$ &         \\
    Schema complement~\citep{DasSarma:2012:FRT}                     &         &         & $\surd$ & $\surd$ &         \\
    Entity complement~\citep{DasSarma:2012:FRT}                    &         &         & $\surd$ &         &         \\
    \citet{Nguyen:2015:RSS}                        &         &         &         & $\surd$ & $\surd$ \\
    InfoGather~\citep{Yakout:2012:IEA}                   &         & $\surd$ &         & $\surd$ & $\surd$ \\
    \bottomrule
  \end{tabular}
  \label{tbl:baseline_method}
\end{table}
}

\vspace*{-0.5\baselineskip}
\subsection{Experimental Setup}
\label{sec:expsetup}
 
The experimental configurations of the various methods are as follows.
For \emph{keyword-based search}, the $T_E$ and $T_H$ methods query an index of the table corpus against the respective fields, while the $T_c$ variant searches against both the caption and catchall fields; all the three methods use BM25. For the \emph{Mannheim Search Join Engine}, the edit distance threshold is set to $\delta=0.8$. For \emph{schema complement}, the heading frequency statistics is calculated based on the Wikipedia table corpora and the heading similarity is aggregated using average. For \emph{entity complement}, WLM is based on entity out-links.  The data similarity threshold is set the same  as for string comparison, i.e., $\delta=0.8$. To parse terms in attribute values, we remove stopwords and HTML markup, and lowercase tokens.  For \emph{Nguyen et al.}, the smoothing parameter value is taken from~\citep{Nguyen:2015:RSS} to be $\alpha=0.5$.  \emph{InfoGather} is trained using linear regression with coordinate ascent. All methods introduced by us, i.e., \emph{HCF-X} and \emph{CRAB-X}, are trained using Random Forest Regression with 5-fold cross-validation; the number of trees is 1000 and the maximum number features is 3.

Table~\ref{tbl:results} presents the evaluation results for the eight baselines.
Among the three keyword-based search methods, which operate on a single table element (top 3 lines), the one that uses table headings as the keyword query performs the best, followed by table entities and table caption.
The methods in lines 4--8 consider multiple table elements; all of these outperform the best single-element method.  The approach that performs best among all, by a large margin, is InfoGather, which incorporates four different table elements.
Consequently, we will test our methods against InfoGather.

\begin{table}[t]
  \centering
  \caption{Evaluation results for existing methods from the literature. Best scores for each metric are boldfaced.}
  \begin{tabular}{ll@{~~}l@{~~}}
    \toprule 
	\textbf{Method} & \small{\textbf{NDCG@5}} & \small{\textbf{NDCG@10}} \\
	\midrule
	Keyword-based search using $T_E$  & 0.2001 & 0.1998 \\
	Keyword-based search using $T_H$  & 0.2318 & 0.2527 \\
	Keyword-based search using $T_c$  & 0.1369 & 0.1419 \\
	Mannheim Search Join Engine  & 0.3298 & 0.3131 \\
	Schema complement  & 0.3389 & 0.3418 \\
	Entity complement  & 0.2986 & 0.3093 \\
	Nguyen et al.  & 0.2875 & 0.3007\\
	InfoGather  & \textbf{0.4530} & \textbf{0.4686} \\	
    \bottomrule
  \end{tabular}
  \label{tbl:results}
\end{table}
%

\subsection{Results}
\label{sub:erom}

\begin{figure}[tbp]
   \centering
   \includegraphics[width=0.46\textwidth]{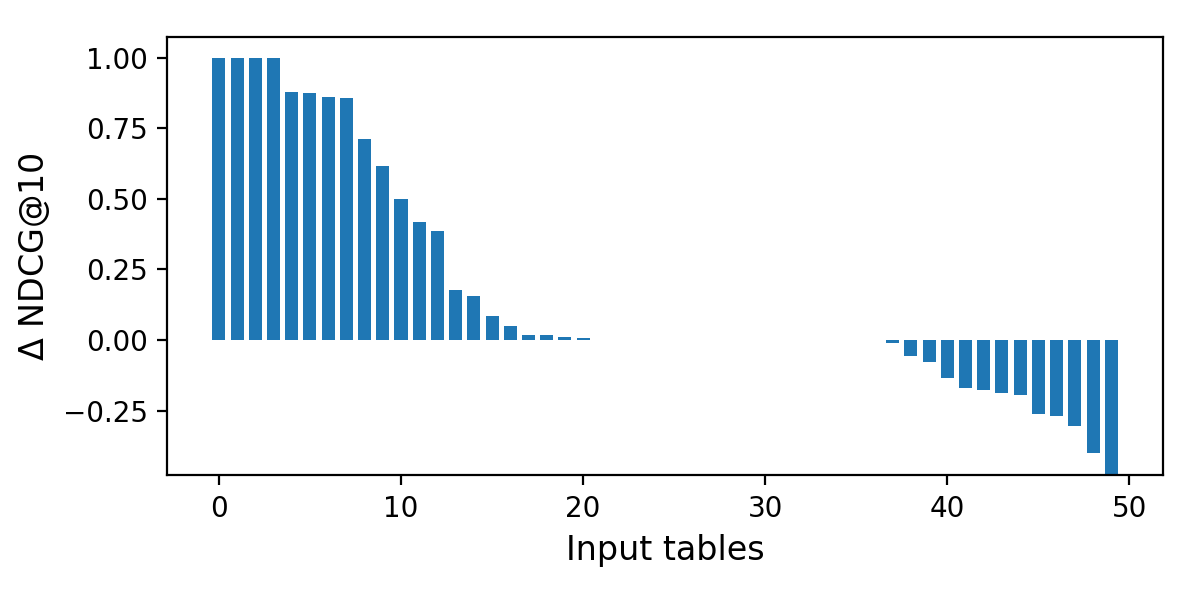}   
   \vspace*{-0.5\baselineskip}
   \caption{Performance difference between InfoGather (baseline) and CRAB-2 on the level of individual input tables. Positive bars indicate substantial advantage of CRAB-2.}
   \vspace*{-0.5\baselineskip}
\label{fig:f_tlevel}
\end{figure}

Table~\ref{tbl:results_crab} compares the evaluation results of the methods we developed in this paper against InfoGather.
HCF-1, which combines all table similarity features from existing approaches, achieves 18.27\% improvement upon InfoGather in terms of NDCG@10, albeit the differences are not statistically significant.
HCF-2 incorporates additional table features, which leads to substantial (29.11\% for NDCG@10) and significant improvements over InfoGather.
The bottom block of Table~\ref{tbl:results_crab} presents the evaluation results for four specific instantiations of our table matching framework (cf. Table~\ref{tbl:crab}). 
Recall that CRAB-1 employs only table similarity features, thus it is to be compared against HCF-1.
CRAB-2..4 additionally consider table features, which corresponds to the settings in HCF-2.
We find that CRAB-1 and CRAB-2 outperform the respective HCF method, while CRAB-4 is on par with it.  None of the differences between CRAB-X and the respective HCF method are statistically significant.
The best overall performer is CRAB-2, with a relative improvement of 36.2\% for NDCG@5 and 33.7\% for NDCG@10 over InfoGather.
Figure~\ref{fig:f_tlevel} shows performance differences on the level of individual input tables between InfoGather and CRAB-2.  Clearly, several tables are improved by a large margin, while only a handful of tables are affected negatively.

The summary of our findings thus far is that our semantic table element representation with element-wise matching is very effective.  We can achieve the same performance as a state-of-the-art approach that relies on hand-crafted features (CRAB-1 vs. HCF-1 and CRAB-2 vs. HCF-2).  
With that, we have accomplished our main research objective.
We further observe that cross-element matching is less effective than element-wise matching (CRAB-3 vs. CRAB-2).  Combining all element-wise and cross-element features performs worse than using only the former (CRAB-4 vs. CRAB-2).

Now that we have assessed the overall effectiveness of our approach, let us turn to answering a series of more specific research questions.

\begin{table}[t]
  \centering
  \caption{Evaluation of our table recommendation methods against the best existing method. Significance is tested against InfoGather. Highest scores are in boldface.}
  \begin{tabular}{ l  ll }
    \toprule 
	\textbf{Method} & \textbf{NDCG@5} & \textbf{NDCG@10}  \\
	\midrule
	InfoGather & 0.4530 & 0.4686 \\	
	\midrule	
	HCF-1 (feats. from Table~\ref{tbl:features_tsf}) & 0.5382 & 0.5542 \\ 
	HCF-2 (feats. from Tables~\ref{tbl:features_tsf} and \ref{tbl:features_table}) & \textbf{0.5895$\dag$} & \textbf{0.6050$\dag$} 	\\
	\midrule
	CRAB-1 & 0.5578 & 0.5672 \\
	CRAB-2 & \textbf{0.6172$\ddag$} & \textbf{0.6267$\ddag$} \\
	CRAB-3 & 0.5140 & 0.5282 \\	
	CRAB-4 & 0.5804$\dag$ & 0.6027$\dag$ \\	
    \bottomrule
  \end{tabular}
  \label{tbl:results_crab}
\end{table}



\begin{table*}[!th]
\caption{Element-wise similarities for various semantic representations.  Rows and columns corresponds to elements of the input and candidate tables, respectively. The evaluation metric is NDCG@10.  The best scores for each row are in boldface.}
\label{tbl:results_ewce}

\begin{tabular}{llll l llll l llll}
    \toprule 
  	\multicolumn{4}{c}{Word-based} & &  
  	\multicolumn{4}{c}{Graph-based} & &   	
  	\multicolumn{4}{c}{Entity-based} \\
  	\cline{1-4} \cline{6-9} \cline{11-14}
	 & \textbf{$T_{t}$} & \textbf{$T_{H}$} & \textbf{$T_{D}$} & & 
	 & \textbf{$T_{t}$} & \textbf{$T_{E}$} & \textbf{$T_{D}$} & & 
	 & \textbf{$T_{t}$} & \textbf{$T_{E}$} & \textbf{$T_{D}$}  \\
  	\cline{2-4} \cline{7-9} \cline{12-14}
	\textbf{$\tilde{T}_{t}$} & \textbf{0.2814} & 0.0261  & 0.0436 & &
	\textbf{$\tilde{T}_{t}$} & \textbf{0.2765} & 0.0546 & 0.0430 & & 
	\textbf{$\tilde{T}_{t}$} & \textbf{0.4796} & 0.0808 & 0.0644  \\
	\textbf{$\tilde{T}_{H}$} & 0.0336 & \textbf{0.1694} & 0.0288 & &
	\textbf{$\tilde{T}_{E}$} & \textbf{0.0700}  & 0.0679 & 0.0501 & &
	\textbf{$\tilde{T}_{E}$} & 0.0705 & 0.0617 & \textbf{0.0725}  \\
	\textbf{$\tilde{T}_{D}$} & 0.0509 & 0.0183 & \textbf{0.1276} & &
	\textbf{$\tilde{T}_{D}$} & \textbf{0.1012} & 0.0423 & 0.0259 & &
	\textbf{$\tilde{T}_{D}$} & \textbf{0.1052} & 0.0812 & 0.0610  \\
    \bottomrule
  \end{tabular}
 
\end{table*}

\begin{description}
	\item[RQ1] Which of the semantic representations (word-based, graph-based, or entity-based) is the most effective for modeling table elements? 
\end{description}

\noindent
Table~\ref{tbl:results_repr} displays results for each of the three semantic representations.  Among those, entity-based performs the best, followed by word-based and graph-based.  
The differences between entity-based and word-based are significant ($p<$ 0.01), but not between the other pairs of representations.
Interestingly, the entity-based representation delivers performance that is comparable to that of the best existing method, InfoGather (cf. Table~\ref{tbl:results}). 
When combing all three semantic representations (line 4, which is the the same as CRAB-1 in Table~\ref{tbl:results_crab}), we obtain substantial and significant improvements ($p<$0.01) over each individual representation.  This shows the complimentary nature of these semantic representations.

\begin{table}[t]
  \centering
  \caption{Comparison of semantic representations. The significance of combined method is tested against entity-based method.}
  \begin{tabular}{ l  ll }
    \toprule 
	\textbf{Semantic Repr.} & \textbf{NDCG@5} & \textbf{NDCG@10}  \\
	\midrule
	Word-based & 0.3779 & 0.3906 \\
	Graph-based & 0.3012 & 0.3376 \\
	Entity-based & 0.4484 & 0.4884 \\
	\midrule
	Combined & \textbf{0.5578$\dag$} & \textbf{0.5672$\dag$} \\
    \bottomrule
  \end{tabular}
  \label{tbl:results_repr}
\end{table}

\begin{description}
	\item[RQ2] Which of the two element-level matching strategies performs better, ele\-ment-wise or cross-element?
\end{description}

\noindent
We found that adding all the cross-element similarities hurts (CRAB-4 vs. CRAB-2 in Table~\ref{tbl:results_crab}).  In order to get a better understanding of how the element-wise and cross-element matching strategies compare against each other, we break down recommendation performance for all table element pairs according to the different semantic representations in Table~\ref{tbl:results_ewce}.  That is, we rank tables by measuring similarity only between that pair of elements (4 table similarity features in total).
Here, diagonal cells correspond to element-wise matching and all other cells correspond to cross-element matching.  We observe that element-wise matching works best across the board.  This is in line with our earlier findings, i.e., CRAB-2 vs. CRAB-3 in Table~\ref{tbl:results_crab}.
However, for graph-based and entity-based representations, there are several cases where cross-element matching yields higher scores than element-wise matching.  Notably, input table data ($\tilde{T}_D$) has much higher similarity against the topic of the candidate table ($T_t$) than against its data ($T_D$) element, for both graph-based and entity-based representations. This shows that cross-element matching does have merit for certain table element pairs. We perform further analysis in Sect.~\ref{sec:eval:analysis_fa}.

\begin{description}
	\item[RQ3] How much do different table elements contribute to recommendation performance? 
\end{description}

\noindent
To explore the importance of table elements, we turn to Table~\ref{tbl:results_ewce} once again.
We first compare the results for element-wise similarity (i.e., the diagonals) and find that among the four table elements, table topic ($\tilde{T}_{t}\leftrightarrow T_{t}$) contributes the most and table data ($\tilde{T}_{D}\leftrightarrow T_{D}$) contributes the least.
Second, our observations for cross-element matching are as follows. Using word-based representation, table data ($\tilde{T}_D$) is the most important element for the input table, while for the candidate table it is table topic ($T_t$).  Interestingly, for graph-based and entity-based representations it is exactly the other way around: the most important input table element is topic ($\tilde{T}_t$), while the most important candidate table element is data ($T_D$).

\subsection{Further Analysis}
\label{sec:eval:analysis}

Now that we have represented our experimental results, we perform further performance analysis on individual features and on input table size.

\subsubsection{Feature Analysis}
\label{sec:eval:analysis_fa}

\begin{figure}[tbp]
   \centering
   \includegraphics[width=0.45\textwidth]{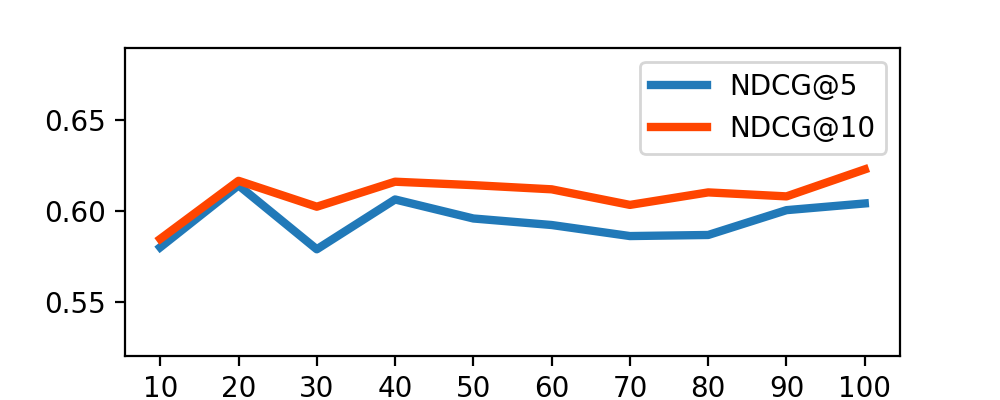} 
   \caption{Performance in terms of NDCG with different number of top features utilized.}
   \vspace*{-1\baselineskip}
\label{fig:top_k_feature}
\end{figure}
\begin{figure}[tbp]
   \centering
   \includegraphics[width=0.46\textwidth]{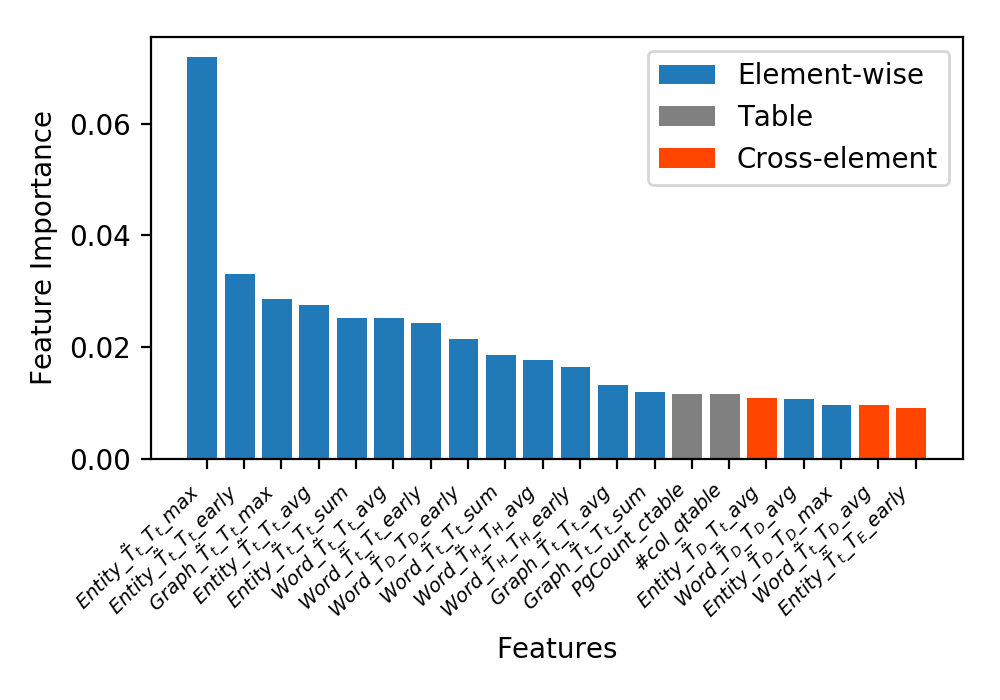}   
   \vspace*{-\baselineskip}
   \caption{Top-20 features in terms of Gini importance.}
   \vspace*{-0.5\baselineskip}
\label{fig:f_imp}
\end{figure}

To understand the contributions of individual features, we first rank all features based on Gini importance. 
Then, we incrementally add features in batches of 10, and plot the corresponding recommendation performance in Figure~\ref{fig:top_k_feature}.
We observe that we can reach peak performance with using only the top-20 features.
%
Let us take a closer look at those top-20 features in Figure~\ref{fig:f_imp}.
We use color coding to help distinguish between the three main types of features: element-wise, cross-element, and table features.  Then, based on these feature importance scores, we revisit our research questions.
Concerning semantic representations (\textbf{RQ1}), there are 8 word embedding, 7 entity embedding, and 3 graph embedding features in the top 20.  Even though there are slightly more features using word embedding than entity embeddings, the latter features are much higher ranked (cf. Fig.~\ref{fig:f_imp}).  Thus, the entity-based semantic representation is the most effective one.
Comparing matching strategies (\textbf{RQ2}), the numbers of element-wise and cross-wise features are 15 and 3, respectively.  This indicates a substantial advantage of element-wise strategies.  Nevertheless, it shows that incorporating the similarity between elements of different types can also be beneficial.  Additionally, there are 2 table features in the top 20.  
As for the importance of table elements (\textbf{RQ3}), table topic ($T_{t}$) is clearly the most important one; 8 out of the top 10 features consider that element.
In summary, our observations based on the top-20 features are consistent with our earlier findings. 


%
%

\subsubsection{Input table size}

\begin{figure}[tbp]
   \centering
   \includegraphics[width=0.45\textwidth]{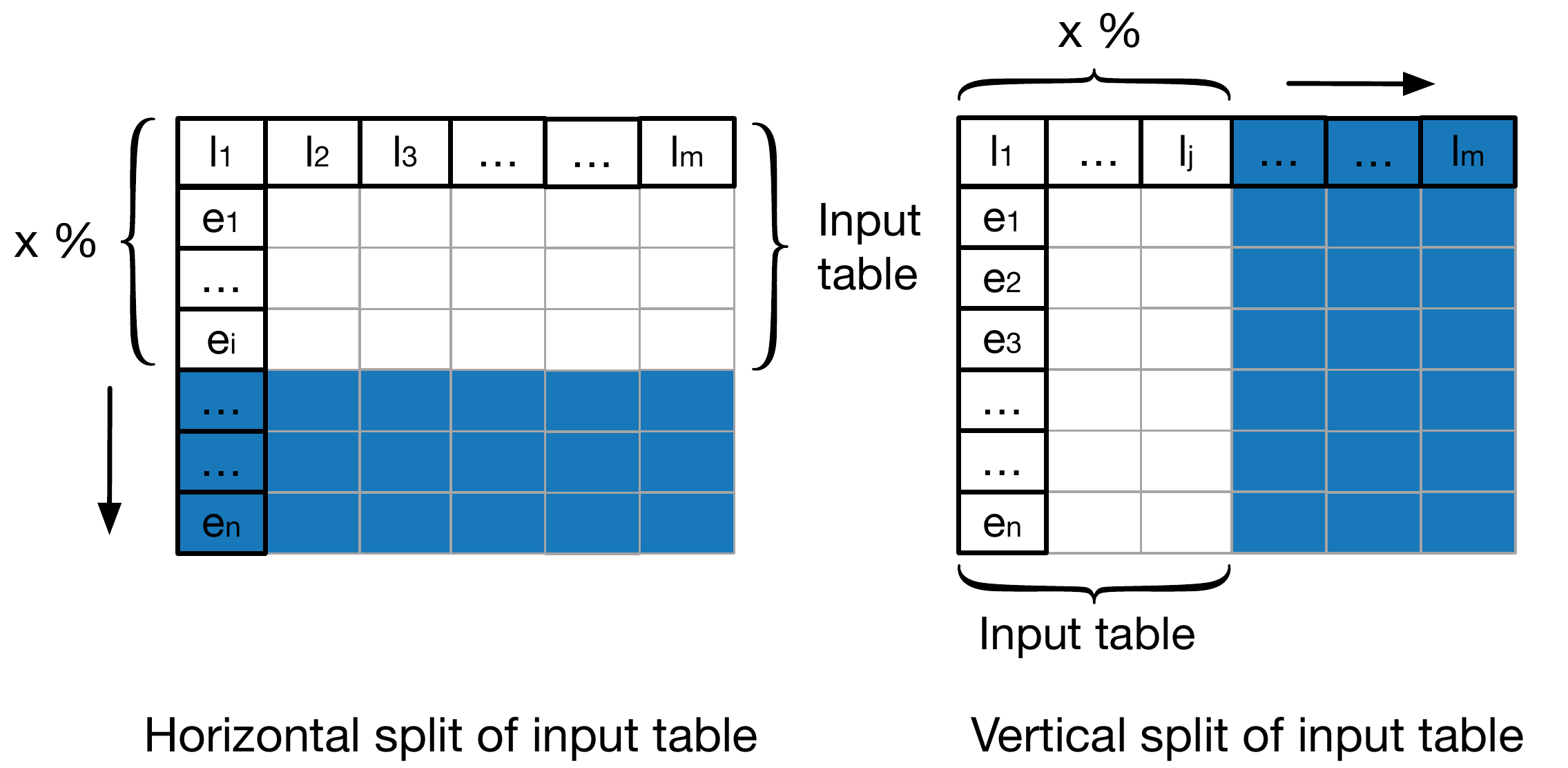}  
   \caption{Performance analysis using only a portion of the table as input.}
      \vspace*{-0.5\baselineskip}
   \label{fig:query_split}
\end{figure}

\begin{figure}[tbp]
    \begin{subfigure}[b]{0.21\textwidth}
        \centering
        \includegraphics[width = 1\textwidth]{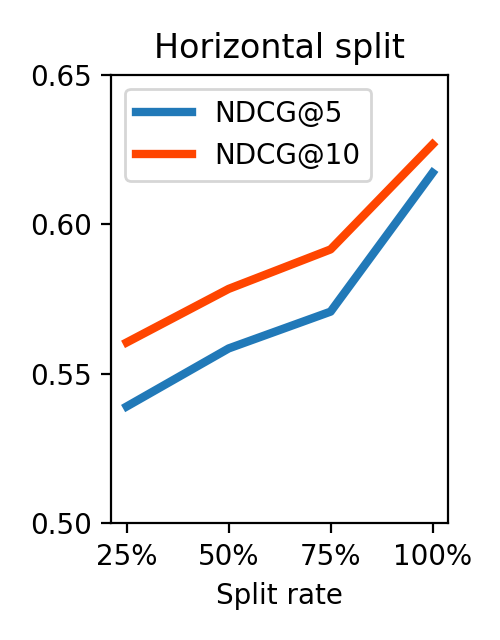}
    \end{subfigure}
    \begin{subfigure}[b]{0.21\textwidth}
        \centering
        \includegraphics[width = 1\textwidth]{figures/horizontal_split.png}       
    \end{subfigure}
    \caption{Performance of CRAB-2 with respect to (relative) input table size, by varying the number of rows (Left) or columns (Right).}
\label{fig:horizontal_split}
\end{figure}

Next, we explore how the size of the input table affects recommendation performance. 
Specifically, we vary the input table size by splitting it horizontally (varying the number of rows) or vertically (varying the number of columns), and using only a portion of the table as input; see Fig.~\ref{fig:query_split} for an illustration.  We explore four settings by setting the split rate $x$ between 25\% and 100\% in steps of  25\%.
Figure~\ref{fig:horizontal_split} plots recommendation performance against input table size.
We observe that growing the table, either horizontally or vertically, results in proportional increase in recommendation performance.  This is not surprising, given that larger tables contain more information.  Nevertheless, being able to utilize this extra information effectively is an essential characteristic of our table matching framework.

\section{Related Work}
An increasing number of studies on web tables are addressing various table-related tasks, including table search, table augmentation, table mining, etc. Among them, table search is considered as a fundamental task both on its own and as a component in other tasks.
%
\emph{Table search} answers a query with a ranked list of tables. 
Early work solves this task for keyword queries~\citep{Cafarella:2008:WEP,Cafarella:2009:DIR,Venetis:2011:RST,Pimplikar:2012:ATQ,Balakrishnan:2015:AWP,Nguyen:2015:RSS}. The WebTables system by \citet{Cafarella:2008:WEP} pioneered keyword-based  table search by issuing the query to a search engine and filtering tables from the returned web pages; the same approach is implemented in \cite{Cafarella:2009:DIR} as well. 
\citet{Venetis:2011:RST} utilize a database of class labels and relationships extracted from the Web for table search.
Using column keywords, \citet{Pimplikar:2012:ATQ} search tables using the term matches in the header, body and context of tables as signals. An example of a keyword-based table search system interface is provided by Google Web Tables.\footnote{https://research.google.com/tables} The developers of this system summarize their experiences in~\cite{Balakrishnan:2015:AWP}. To enrich the diversity of search results,~\citet{Nguyen:2015:RSS} design a goodness measure for table search and selection.  Their query is not limited to keywords, it can also be a table.  
We have discussed the line of approaches~\cite{Lehmberg:2015:MSJ, Ahmadov:2015:THI, DasSarma:2012:FRT, Yakout:2012:IEA, Nguyen:2015:RSS, Limaye:2010:ASW} that can use a table as a query in Sect.~\ref{sub:er}.

%
%
%

%

Tables contain rich knowledge and have raised great interest in \emph{table mining} research~\cite{Cafarella:2011:SDW,Cafarella:2008:WEP,JM:2009:HDW,Sarawagi:2014:OQQ,Venetis:2011:RST,Zhang:2013:ISM, Arvind:2015:NPI, Yin:2016:NEL, Sarawagi:2014:OQQ,Banerjee:2009:LRQ, Venetis:2011:RST, Chirigati:2016:KEU, Bhagavatula:2015:TEL}. \citet{Munoz:2014:ULD} recover Wikipedia table semantics and store them in RDF triples. Similar work is taken in \cite{Cafarella:2008:WEP} based on tables extracted from a Google crawl. Instead of mining a whole table corpora, a single table store many facts, which could be answers for questions. \citet{Yin:2016:NEL} take a single table as a knowledge base and perform querying on it using deep neural networks. The knowledge extracted from tables could be used to augment an existing knowledge base~\cite{Sekhavat:2014:KBA, Dong:2014:KVW}. E.g., \citet{Sekhavat:2014:KBA} and \citet{Dong:2014:KVW} design probabilistic methods to utilize table information for  augmenting an existing knowledge base.  Another line of work concerns table annotation and classification. By mining column content, \citet{Zwicklbauer:2013:TDW} propose a method to annotate table headers. Studying a large number of tables in \cite{Crestan:2011:WTC}, a well defined table type taxonomy is provided for classifying HTML tables. Besides all the above tasks, table mining refers to tasks like \emph{table interpretation}~\cite{Cafarella:2008:WEP, Munoz:2014:ULD, Venetis:2011:RST} and \emph{table recognition}~\cite{Crestan:2011:WTC,Zwicklbauer:2013:TDW} as well. 


\emph{Table augmentation} is the task of extending a table with additional elements, e.g., new columns~\cite{DasSarma:2012:FRT,Cafarella:2009:DIR,Lehmberg:2015:MSJ,Yakout:2012:IEA,Bhagavatula:2013:MEM}. To capture relevant data, e.g., existing columns, these methods need to search tables~\cite{Lehmberg:2015:MSJ,Bhagavatula:2013:MEM,Yakout:2012:IEA}. E.g., the Mannheim Search Join Engine~\cite{Lehmberg:2015:MSJ} searches the top-$k$ candidate tables from a corpus of web tables and picks relevant columns to merge. Extending a table with more rows also needs table retrieval~\cite{DasSarma:2012:FRT,Yakout:2012:IEA,Zhang:2017:ESA, Zhang:2017:ESA}. In~\cite{Zhang:2017:ESA}, two tasks of row population and column population are proposed, which provide suggestions for extending an entity-focused table with additional rows and columns. \emph{Table completion} is the task of filling in empty cells within a table. \citet{Ahmadov:2015:THI} introduce a method to extract table values from related tables and/or to predict them using machine learning methods.


\section{Conclusions}
In this paper, we have introduced and addressed the task of \emph{recommending related tables}: returning a ranked list of tables that are related to a given input table.
We have proposed a novel element-oriented table matching framework that represents table elements uniformly and considers their similarity in multiple semantic spaces.  
This framework can incorporate the similarity between table elements that are of the same type (element-wise matching) as well as those that are of different types (cross-element matching).
We have further presented four specific instantiations of this general framework and considered word-based, graph-based, and entity-based semantic representations.
For evaluation, we have developed a standard test collection based on Wikipedia tables, and demonstrated that our approach delivers state-of-the-art performance.

In the future, we plan to test our method on a more heterogeneous collection of tables from the Web, which vary more quality-wise than Wikipedia tables.  We are further interested in evaluating the utility of our approach with user studies.


\balance
\bibliographystyle{ACM-Reference-Format}
\bibliography{00paper}


\begin{thebibliography}{44}


\ifx \showCODEN    \undefined \def \showCODEN     #1{\unskip}     \fi
\ifx \showDOI      \undefined \def \showDOI       #1{#1}\fi
\ifx \showISBNx    \undefined \def \showISBNx     #1{\unskip}     \fi
\ifx \showISBNxiii \undefined \def \showISBNxiii  #1{\unskip}     \fi
\ifx \showISSN     \undefined \def \showISSN      #1{\unskip}     \fi
\ifx \showLCCN     \undefined \def \showLCCN      #1{\unskip}     \fi
\ifx \shownote     \undefined \def \shownote      #1{#1}          \fi
\ifx \showarticletitle \undefined \def \showarticletitle #1{#1}   \fi
\ifx \showURL      \undefined \def \showURL       {\relax}        \fi
\providecommand\bibfield[2]{#2}
\providecommand\bibinfo[2]{#2}
\providecommand\natexlab[1]{#1}
\providecommand\showeprint[2][]{arXiv:#2}

\bibitem[\protect\citeauthoryear{Ahmadov, Thiele, Eberius, Lehner, and
  Wrembel}{Ahmadov et~al\mbox{.}}{2015}]%
        {Ahmadov:2015:THI}
\bibfield{author}{\bibinfo{person}{Ahmad Ahmadov}, \bibinfo{person}{Maik
  Thiele}, \bibinfo{person}{Julian Eberius}, \bibinfo{person}{Wolfgang Lehner},
  {and} \bibinfo{person}{Robert Wrembel}.} \bibinfo{year}{2015}\natexlab{}.
\newblock \showarticletitle{Towards a Hybrid Imputation Approach Using Web
  Tables.}. In \bibinfo{booktitle}{\emph{Proc. of BDC '15}}.
  \bibinfo{pages}{21--30}.
\newblock


\bibitem[\protect\citeauthoryear{Anonymous}{Anonymous}{2017}]%
        {Zhang:2017:ATR}
\bibfield{author}{\bibinfo{person}{Anonymous}.}
  \bibinfo{year}{2017}\natexlab{}.
\newblock \showarticletitle{Removed to Protect Anonymity}.
\newblock  (\bibinfo{year}{2017}).
\newblock


\bibitem[\protect\citeauthoryear{Balakrishnan, Halevy, Harb, Lee, Madhavan,
  Rostamizadeh, Shen, Wilder, Wu, and Yu}{Balakrishnan et~al\mbox{.}}{2015}]%
        {Balakrishnan:2015:AWP}
\bibfield{author}{\bibinfo{person}{Sreeram Balakrishnan},
  \bibinfo{person}{Alon~Y. Halevy}, \bibinfo{person}{Boulos Harb},
  \bibinfo{person}{Hongrae Lee}, \bibinfo{person}{Jayant Madhavan},
  \bibinfo{person}{Afshin Rostamizadeh}, \bibinfo{person}{Warren Shen},
  \bibinfo{person}{Kenneth Wilder}, \bibinfo{person}{Fei Wu}, {and}
  \bibinfo{person}{Cong Yu}.} \bibinfo{year}{2015}\natexlab{}.
\newblock \showarticletitle{Applying WebTables in Practice}. In
  \bibinfo{booktitle}{\emph{Proc. of CIDR '15}}.
\newblock


\bibitem[\protect\citeauthoryear{Banerjee, Chakrabarti, and
  Ramakrishnan}{Banerjee et~al\mbox{.}}{2009}]%
        {Banerjee:2009:LRQ}
\bibfield{author}{\bibinfo{person}{Somnath Banerjee}, \bibinfo{person}{Soumen
  Chakrabarti}, {and} \bibinfo{person}{Ganesh Ramakrishnan}.}
  \bibinfo{year}{2009}\natexlab{}.
\newblock \showarticletitle{Learning to Rank for Quantity Consensus Queries}.
  In \bibinfo{booktitle}{\emph{Proc. of SIGIR '09}}. \bibinfo{pages}{243--250}.
\newblock


\bibitem[\protect\citeauthoryear{Bhagavatula, Noraset, and Downey}{Bhagavatula
  et~al\mbox{.}}{2013}]%
        {Bhagavatula:2013:MEM}
\bibfield{author}{\bibinfo{person}{Chandra~Sekhar Bhagavatula},
  \bibinfo{person}{Thanapon Noraset}, {and} \bibinfo{person}{Doug Downey}.}
  \bibinfo{year}{2013}\natexlab{}.
\newblock \showarticletitle{Methods for Exploring and Mining Tables on
  Wikipedia}. In \bibinfo{booktitle}{\emph{Proc. of IDEA '13}}.
  \bibinfo{pages}{18--26}.
\newblock


\bibitem[\protect\citeauthoryear{Bhagavatula, Noraset, and Downey}{Bhagavatula
  et~al\mbox{.}}{2015}]%
        {Bhagavatula:2015:TEL}
\bibfield{author}{\bibinfo{person}{Chandra~Sekhar Bhagavatula},
  \bibinfo{person}{Thanapon Noraset}, {and} \bibinfo{person}{Doug Downey}.}
  \bibinfo{year}{2015}\natexlab{}.
\newblock \showarticletitle{TabEL: Entity Linking in Web Tables}. In
  \bibinfo{booktitle}{\emph{Proc. of ISWC 2015}}. \bibinfo{pages}{425--441}.
\newblock


\bibitem[\protect\citeauthoryear{Cafarella, Halevy, and Khoussainova}{Cafarella
  et~al\mbox{.}}{2009}]%
        {Cafarella:2009:DIR}
\bibfield{author}{\bibinfo{person}{Michael~J. Cafarella}, \bibinfo{person}{Alon
  Halevy}, {and} \bibinfo{person}{Nodira Khoussainova}.}
  \bibinfo{year}{2009}\natexlab{}.
\newblock \showarticletitle{Data Integration for the Relational Web}.
\newblock \bibinfo{journal}{\emph{Proc. of VLDB Endow.}}  \bibinfo{volume}{2}
  (\bibinfo{year}{2009}), \bibinfo{pages}{1090--1101}.
\newblock


\bibitem[\protect\citeauthoryear{Cafarella, Halevy, and Madhavan}{Cafarella
  et~al\mbox{.}}{2011}]%
        {Cafarella:2011:SDW}
\bibfield{author}{\bibinfo{person}{Michael~J. Cafarella}, \bibinfo{person}{Alon
  Halevy}, {and} \bibinfo{person}{Jayant Madhavan}.}
  \bibinfo{year}{2011}\natexlab{}.
\newblock \showarticletitle{Structured Data on the Web}.
\newblock \bibinfo{journal}{\emph{Commun. ACM}}  \bibinfo{volume}{54}
  (\bibinfo{year}{2011}), \bibinfo{pages}{72--79}.
\newblock


\bibitem[\protect\citeauthoryear{Cafarella, Halevy, Wang, Wu, and
  Zhang}{Cafarella et~al\mbox{.}}{2008}]%
        {Cafarella:2008:WEP}
\bibfield{author}{\bibinfo{person}{Michael~J. Cafarella}, \bibinfo{person}{Alon
  Halevy}, \bibinfo{person}{Daisy~Zhe Wang}, \bibinfo{person}{Eugene Wu}, {and}
  \bibinfo{person}{Yang Zhang}.} \bibinfo{year}{2008}\natexlab{}.
\newblock \showarticletitle{WebTables: Exploring the Power of Tables on the
  Web}.
\newblock \bibinfo{journal}{\emph{Proc. of VLDB Endow.}}  \bibinfo{volume}{1}
  (\bibinfo{year}{2008}), \bibinfo{pages}{538--549}.
\newblock


\bibitem[\protect\citeauthoryear{Chirigati, Liu, Korn, Wu, Yu, and
  Zhang}{Chirigati et~al\mbox{.}}{2016}]%
        {Chirigati:2016:KEU}
\bibfield{author}{\bibinfo{person}{Fernando Chirigati}, \bibinfo{person}{Jialu
  Liu}, \bibinfo{person}{Flip Korn}, \bibinfo{person}{You~(Will) Wu},
  \bibinfo{person}{Cong Yu}, {and} \bibinfo{person}{Hao Zhang}.}
  \bibinfo{year}{2016}\natexlab{}.
\newblock \showarticletitle{Knowledge Exploration Using Tables on the Web}.
\newblock \bibinfo{journal}{\emph{Proc. of VLDB Endow.}}  \bibinfo{volume}{10}
  (\bibinfo{year}{2016}), \bibinfo{pages}{193--204}.
\newblock


\bibitem[\protect\citeauthoryear{Crestan and Pantel}{Crestan and
  Pantel}{2011}]%
        {Crestan:2011:WTC}
\bibfield{author}{\bibinfo{person}{Eric Crestan} {and} \bibinfo{person}{Patrick
  Pantel}.} \bibinfo{year}{2011}\natexlab{}.
\newblock \showarticletitle{Web-scale Table Census and Classification}. In
  \bibinfo{booktitle}{\emph{Proc. of WSDM '11}}. \bibinfo{pages}{545--554}.
\newblock


\bibitem[\protect\citeauthoryear{Das~Sarma, Fang, Gupta, Halevy, Lee, Wu, Xin,
  and Yu}{Das~Sarma et~al\mbox{.}}{2012}]%
        {DasSarma:2012:FRT}
\bibfield{author}{\bibinfo{person}{Anish Das~Sarma}, \bibinfo{person}{Lujun
  Fang}, \bibinfo{person}{Nitin Gupta}, \bibinfo{person}{Alon Halevy},
  \bibinfo{person}{Hongrae Lee}, \bibinfo{person}{Fei Wu},
  \bibinfo{person}{Reynold Xin}, {and} \bibinfo{person}{Cong Yu}.}
  \bibinfo{year}{2012}\natexlab{}.
\newblock \showarticletitle{Finding Related Tables}. In
  \bibinfo{booktitle}{\emph{Proc. of SIGMOD '12}}. \bibinfo{pages}{817--828}.
\newblock


\bibitem[\protect\citeauthoryear{Dong, Gabrilovich, Heitz, Horn, Lao, Murphy,
  Strohmann, Sun, and Zhang}{Dong et~al\mbox{.}}{2014}]%
        {Dong:2014:KVW}
\bibfield{author}{\bibinfo{person}{Xin Dong}, \bibinfo{person}{Evgeniy
  Gabrilovich}, \bibinfo{person}{Geremy Heitz}, \bibinfo{person}{Wilko Horn},
  \bibinfo{person}{Ni Lao}, \bibinfo{person}{Kevin Murphy},
  \bibinfo{person}{Thomas Strohmann}, \bibinfo{person}{Shaohua Sun}, {and}
  \bibinfo{person}{Wei Zhang}.} \bibinfo{year}{2014}\natexlab{}.
\newblock \showarticletitle{Knowledge Vault: A Web-scale Approach to
  Probabilistic Knowledge Fusion}. In \bibinfo{booktitle}{\emph{Proc. of KDD
  '14}}. \bibinfo{pages}{601--610}.
\newblock


\bibitem[\protect\citeauthoryear{Fleiss et~al\mbox{.}}{Fleiss
  et~al\mbox{.}}{1971}]%
        {Fleiss:1971:MNS}
\bibfield{author}{\bibinfo{person}{J.L. Fleiss} {et~al\mbox{.}}}
  \bibinfo{year}{1971}\natexlab{}.
\newblock \showarticletitle{{Measuring nominal scale agreement among many
  raters}}.
\newblock \bibinfo{journal}{\emph{Psychological Bulletin}}
  \bibinfo{volume}{76} (\bibinfo{year}{1971}), \bibinfo{pages}{378--382}.
\newblock


\bibitem[\protect\citeauthoryear{Hasibi, Balog, Garigliotti, and Zhang}{Hasibi
  et~al\mbox{.}}{2017}]%
        {Hasibi:2017:NTE}
\bibfield{author}{\bibinfo{person}{Faegheh Hasibi}, \bibinfo{person}{Krisztian
  Balog}, \bibinfo{person}{Dar\'{\i}o Garigliotti}, {and} \bibinfo{person}{Shuo
  Zhang}.} \bibinfo{year}{2017}\natexlab{}.
\newblock \showarticletitle{Nordlys: A Toolkit for Entity-Oriented and Semantic
  Search}. In \bibinfo{booktitle}{\emph{Proc. of SIGIR '17}}.
  \bibinfo{pages}{1289--1292}.
\newblock


\bibitem[\protect\citeauthoryear{Ibrahim, Riedewald, and Weikum}{Ibrahim
  et~al\mbox{.}}{2016}]%
        {Ibrahim:2016:MSE}
\bibfield{author}{\bibinfo{person}{Yusra Ibrahim}, \bibinfo{person}{Mirek
  Riedewald}, {and} \bibinfo{person}{Gerhard Weikum}.}
  \bibinfo{year}{2016}\natexlab{}.
\newblock \showarticletitle{Making Sense of Entities and Quantities in Web
  Tables}. In \bibinfo{booktitle}{\emph{Proc. of CIKM '16}}.
  \bibinfo{pages}{1703--1712}.
\newblock
\showISBNx{978-1-4503-4073-1}


\bibitem[\protect\citeauthoryear{Lehmberg, Ritze, Meusel, and Bizer}{Lehmberg
  et~al\mbox{.}}{2016}]%
        {Lehmberg:2016:LPC}
\bibfield{author}{\bibinfo{person}{Oliver Lehmberg}, \bibinfo{person}{Dominique
  Ritze}, \bibinfo{person}{Robert Meusel}, {and} \bibinfo{person}{Christian
  Bizer}.} \bibinfo{year}{2016}\natexlab{}.
\newblock \showarticletitle{A Large Public Corpus of Web Tables Containing Time
  and Context Metadata}. In \bibinfo{booktitle}{\emph{Proc. of WWW '16
  Companion}}. \bibinfo{pages}{75--76}.
\newblock


\bibitem[\protect\citeauthoryear{Lehmberg, Ritze, Ristoski, Meusel, Paulheim,
  and Bizer}{Lehmberg et~al\mbox{.}}{2015}]%
        {Lehmberg:2015:MSJ}
\bibfield{author}{\bibinfo{person}{Oliver Lehmberg}, \bibinfo{person}{Dominique
  Ritze}, \bibinfo{person}{Petar Ristoski}, \bibinfo{person}{Robert Meusel},
  \bibinfo{person}{Heiko Paulheim}, {and} \bibinfo{person}{Christian Bizer}.}
  \bibinfo{year}{2015}\natexlab{}.
\newblock \showarticletitle{The Mannheim Search Join Engine}.
\newblock \bibinfo{journal}{\emph{Web Semant.}}  \bibinfo{volume}{35}
  (\bibinfo{year}{2015}), \bibinfo{pages}{159--166}.
\newblock


\bibitem[\protect\citeauthoryear{Limaye, Sarawagi, and Chakrabarti}{Limaye
  et~al\mbox{.}}{2010}]%
        {Limaye:2010:ASW}
\bibfield{author}{\bibinfo{person}{Girija Limaye}, \bibinfo{person}{Sunita
  Sarawagi}, {and} \bibinfo{person}{Soumen Chakrabarti}.}
  \bibinfo{year}{2010}\natexlab{}.
\newblock \showarticletitle{Annotating and Searching Web Tables Using Entities,
  Types and Relationships}.
\newblock \bibinfo{journal}{\emph{Proc. of VLDB Endow.}}  \bibinfo{volume}{3}
  (\bibinfo{year}{2010}), \bibinfo{pages}{1338--1347}.
\newblock


\bibitem[\protect\citeauthoryear{Macdonald, Santos, and Ounis}{Macdonald
  et~al\mbox{.}}{2012}]%
        {Macdonald:2012:UQF}
\bibfield{author}{\bibinfo{person}{Craig Macdonald}, \bibinfo{person}{Rodrygo
  L~T Santos}, {and} \bibinfo{person}{Iadh Ounis}.}
  \bibinfo{year}{2012}\natexlab{}.
\newblock \showarticletitle{On the Usefulness of Query Features for Learning to
  Rank}. In \bibinfo{booktitle}{\emph{Proc. of CIKM '12}}.
  \bibinfo{pages}{2559--2562}.
\newblock


\bibitem[\protect\citeauthoryear{Madhavan, Afanasiev, Antova, and
  Halevy}{Madhavan et~al\mbox{.}}{2009}]%
        {JM:2009:HDW}
\bibfield{author}{\bibinfo{person}{Jayant Madhavan}, \bibinfo{person}{Loredana
  Afanasiev}, \bibinfo{person}{Lyublena Antova}, {and} \bibinfo{person}{Alon~Y.
  Halevy}.} \bibinfo{year}{2009}\natexlab{}.
\newblock \showarticletitle{Harnessing the Deep Web: Present and Future}.
\newblock \bibinfo{journal}{\emph{CoRR}}  \bibinfo{volume}{abs/0909.1785}
  (\bibinfo{year}{2009}).
\newblock


\bibitem[\protect\citeauthoryear{Mikolov, Sutskever, Chen, Corrado, and
  Dean}{Mikolov et~al\mbox{.}}{2013}]%
        {Mikolov:2013:DRW}
\bibfield{author}{\bibinfo{person}{Tomas Mikolov}, \bibinfo{person}{Ilya
  Sutskever}, \bibinfo{person}{Kai Chen}, \bibinfo{person}{Greg Corrado}, {and}
  \bibinfo{person}{Jeffrey Dean}.} \bibinfo{year}{2013}\natexlab{}.
\newblock \showarticletitle{Distributed Representations of Words and Phrases
  and Their Compositionality}. In \bibinfo{booktitle}{\emph{Proc. of NIPS
  '13}}. \bibinfo{pages}{3111--3119}.
\newblock


\bibitem[\protect\citeauthoryear{Munoz, Hogan, and Mileo}{Munoz
  et~al\mbox{.}}{2014}]%
        {Munoz:2014:ULD}
\bibfield{author}{\bibinfo{person}{Emir Munoz}, \bibinfo{person}{Aidan Hogan},
  {and} \bibinfo{person}{Alessandra Mileo}.} \bibinfo{year}{2014}\natexlab{}.
\newblock \showarticletitle{Using Linked Data to Mine RDF from Wikipedia's
  Tables}. In \bibinfo{booktitle}{\emph{Proc. of WSDM '14}}.
  \bibinfo{pages}{533--542}.
\newblock


\bibitem[\protect\citeauthoryear{Neelakantan, Le, and Sutskever}{Neelakantan
  et~al\mbox{.}}{2015}]%
        {Arvind:2015:NPI}
\bibfield{author}{\bibinfo{person}{Arvind Neelakantan},
  \bibinfo{person}{Quoc~V. Le}, {and} \bibinfo{person}{Ilya Sutskever}.}
  \bibinfo{year}{2015}\natexlab{}.
\newblock \showarticletitle{Neural Programmer: Inducing Latent Programs with
  Gradient Descent}.
\newblock \bibinfo{journal}{\emph{CoRR}}  \bibinfo{volume}{abs/1511.04834}
  (\bibinfo{year}{2015}).
\newblock


\bibitem[\protect\citeauthoryear{Nguyen, Nguyen, Matthias, and Karl}{Nguyen
  et~al\mbox{.}}{2015}]%
        {Nguyen:2015:RSS}
\bibfield{author}{\bibinfo{person}{Thanh~Tam Nguyen}, \bibinfo{person}{Quoc
  Viet~Hung Nguyen}, \bibinfo{person}{Weidlich Matthias}, {and}
  \bibinfo{person}{Aberer Karl}.} \bibinfo{year}{2015}\natexlab{}.
\newblock \showarticletitle{Result Selection and Summarization for Web Table
  Search}. In \bibinfo{booktitle}{\emph{ISDE '15}}. \bibinfo{pages}{425--441}.
\newblock


\bibitem[\protect\citeauthoryear{Pennington, Socher, and Manning}{Pennington
  et~al\mbox{.}}{2014}]%
        {Pennington:2014:GGV}
\bibfield{author}{\bibinfo{person}{Jeffrey Pennington},
  \bibinfo{person}{Richard Socher}, {and} \bibinfo{person}{Christopher~D
  Manning}.} \bibinfo{year}{2014}\natexlab{}.
\newblock \showarticletitle{{GloVe}: Global Vectors for Word Representation}.
  In \bibinfo{booktitle}{\emph{Proc. of EMNLP '14}}.
  \bibinfo{pages}{1532--1543}.
\newblock


\bibitem[\protect\citeauthoryear{Pimplikar and Sarawagi}{Pimplikar and
  Sarawagi}{2012}]%
        {Pimplikar:2012:ATQ}
\bibfield{author}{\bibinfo{person}{Rakesh Pimplikar} {and}
  \bibinfo{person}{Sunita Sarawagi}.} \bibinfo{year}{2012}\natexlab{}.
\newblock \showarticletitle{Answering Table Queries on the Web Using Column
  Keywords}.
\newblock \bibinfo{journal}{\emph{Proc. of VLDB Endow.}}  \bibinfo{volume}{5}
  (\bibinfo{year}{2012}), \bibinfo{pages}{908--919}.
\newblock


\bibitem[\protect\citeauthoryear{Qin, Liu, Xu, and Li}{Qin
  et~al\mbox{.}}{2010}]%
        {Qin:2010:LBC}
\bibfield{author}{\bibinfo{person}{Tao Qin}, \bibinfo{person}{Tie-Yan Liu},
  \bibinfo{person}{Jun Xu}, {and} \bibinfo{person}{Hang Li}.}
  \bibinfo{year}{2010}\natexlab{}.
\newblock \showarticletitle{{LETOR}: A Benchmark Collection for Research on
  Learning to Rank for Information Retrieval}.
\newblock \bibinfo{journal}{\emph{Inf. Retr.}} \bibinfo{volume}{13},
  \bibinfo{number}{4} (\bibinfo{date}{Aug} \bibinfo{year}{2010}),
  \bibinfo{pages}{346--374}.
\newblock


\bibitem[\protect\citeauthoryear{Ristoski and Paulheim}{Ristoski and
  Paulheim}{2016}]%
        {Ristoski:2016:RGE}
\bibfield{author}{\bibinfo{person}{Petar Ristoski} {and} \bibinfo{person}{Heiko
  Paulheim}.} \bibinfo{year}{2016}\natexlab{}.
\newblock \showarticletitle{{RDF2vec}: {RDF} Graph Embeddings for Data Mining}.
  In \bibinfo{booktitle}{\emph{Proc. of ISWC '16}}. \bibinfo{pages}{498--514}.
\newblock


\bibitem[\protect\citeauthoryear{Ritze, Lehmberg, Oulabi, and Bizer}{Ritze
  et~al\mbox{.}}{2016}]%
        {Ritze:2016:PPW}
\bibfield{author}{\bibinfo{person}{Dominique Ritze}, \bibinfo{person}{Oliver
  Lehmberg}, \bibinfo{person}{Yaser Oulabi}, {and} \bibinfo{person}{Christian
  Bizer}.} \bibinfo{year}{2016}\natexlab{}.
\newblock \showarticletitle{Profiling the Potential of Web Tables for
  Augmenting Cross-domain Knowledge Bases}. In \bibinfo{booktitle}{\emph{Proc.
  of WWW '16}}. \bibinfo{pages}{251--261}.
\newblock


\bibitem[\protect\citeauthoryear{Sarawagi and Chakrabarti}{Sarawagi and
  Chakrabarti}{2014}]%
        {Sarawagi:2014:OQQ}
\bibfield{author}{\bibinfo{person}{Sunita Sarawagi} {and}
  \bibinfo{person}{Soumen Chakrabarti}.} \bibinfo{year}{2014}\natexlab{}.
\newblock \showarticletitle{Open-domain Quantity Queries on Web Tables:
  Annotation, Response, and Consensus Models}. In
  \bibinfo{booktitle}{\emph{Proc. of KDD '14}}. \bibinfo{pages}{711--720}.
\newblock


\bibitem[\protect\citeauthoryear{Sekhavat, Paolo, Barbosa, and
  Merialdo}{Sekhavat et~al\mbox{.}}{2014}]%
        {Sekhavat:2014:KBA}
\bibfield{author}{\bibinfo{person}{Yoones~A. Sekhavat},
  \bibinfo{person}{Francesco~Di Paolo}, \bibinfo{person}{Denilson Barbosa},
  {and} \bibinfo{person}{Paolo Merialdo}.} \bibinfo{year}{2014}\natexlab{}.
\newblock \showarticletitle{Knowledge Base Augmentation using Tabular Data}. In
  \bibinfo{booktitle}{\emph{Proc. of LDOW '14}}.
\newblock


\bibitem[\protect\citeauthoryear{Shen, Wang, and Han}{Shen
  et~al\mbox{.}}{2015}]%
        {Shen:2015:ELK}
\bibfield{author}{\bibinfo{person}{Wei Shen}, \bibinfo{person}{Jianyong Wang},
  {and} \bibinfo{person}{Jiawei Han}.} \bibinfo{year}{2015}\natexlab{}.
\newblock \showarticletitle{Entity Linking with a Knowledge Base: Issues,
  Techniques, and Solutions}.
\newblock \bibinfo{journal}{\emph{IEEE Trans. Knowl. Data Eng.}}
  \bibinfo{volume}{27}, \bibinfo{number}{2} (\bibinfo{date}{feb}
  \bibinfo{year}{2015}), \bibinfo{pages}{443--460}.
\newblock


\bibitem[\protect\citeauthoryear{Snoek, Worring, and Smeulders}{Snoek
  et~al\mbox{.}}{2005}]%
        {Snoek:2005:EVL}
\bibfield{author}{\bibinfo{person}{Cees G.~M. Snoek}, \bibinfo{person}{Marcel
  Worring}, {and} \bibinfo{person}{Arnold W.~M. Smeulders}.}
  \bibinfo{year}{2005}\natexlab{}.
\newblock \showarticletitle{Early Versus Late Fusion in Semantic Video
  Analysis}. In \bibinfo{booktitle}{\emph{Proc. of MULTIMEDIA '05}}.
  \bibinfo{pages}{399--402}.
\newblock


\bibitem[\protect\citeauthoryear{Venetis, Halevy, Madhavan, Pa\c{s}ca, Shen,
  Wu, Miao, and Wu}{Venetis et~al\mbox{.}}{2011}]%
        {Venetis:2011:RST}
\bibfield{author}{\bibinfo{person}{Petros Venetis}, \bibinfo{person}{Alon
  Halevy}, \bibinfo{person}{Jayant Madhavan}, \bibinfo{person}{Marius
  Pa\c{s}ca}, \bibinfo{person}{Warren Shen}, \bibinfo{person}{Fei Wu},
  \bibinfo{person}{Gengxin Miao}, {and} \bibinfo{person}{Chung Wu}.}
  \bibinfo{year}{2011}\natexlab{}.
\newblock \showarticletitle{Recovering Semantics of Tables on the Web}.
\newblock \bibinfo{journal}{\emph{Proc. of VLDB Endow.}}  \bibinfo{volume}{4}
  (\bibinfo{year}{2011}), \bibinfo{pages}{528--538}.
\newblock


\bibitem[\protect\citeauthoryear{Wang, Li, and Fe}{Wang et~al\mbox{.}}{2011}]%
        {Wang:2011:FEM}
\bibfield{author}{\bibinfo{person}{Jiannan Wang}, \bibinfo{person}{Guoliang
  Li}, {and} \bibinfo{person}{Jianhua Fe}.} \bibinfo{year}{2011}\natexlab{}.
\newblock \showarticletitle{Fast-join: An Efficient Method for Fuzzy Token
  Matching Based String Similarity Join}. In \bibinfo{booktitle}{\emph{Proc. of
  ICDE '11}}. \bibinfo{pages}{458--469}.
\newblock


\bibitem[\protect\citeauthoryear{Yakout, Ganjam, Chakrabarti, and
  Chaudhuri}{Yakout et~al\mbox{.}}{2012}]%
        {Yakout:2012:IEA}
\bibfield{author}{\bibinfo{person}{Mohamed Yakout}, \bibinfo{person}{Kris
  Ganjam}, \bibinfo{person}{Kaushik Chakrabarti}, {and}
  \bibinfo{person}{Surajit Chaudhuri}.} \bibinfo{year}{2012}\natexlab{}.
\newblock \showarticletitle{InfoGather: Entity Augmentation and Attribute
  Discovery by Holistic Matching with Web Tables}. In
  \bibinfo{booktitle}{\emph{Proc. of SIGMOD '12}}. \bibinfo{pages}{97--108}.
\newblock


\bibitem[\protect\citeauthoryear{Yin, Lu, Li, and Kao}{Yin
  et~al\mbox{.}}{2016}]%
        {Yin:2016:NEL}
\bibfield{author}{\bibinfo{person}{Pengcheng Yin}, \bibinfo{person}{Zhengdong
  Lu}, \bibinfo{person}{Hang Li}, {and} \bibinfo{person}{Ben Kao}.}
  \bibinfo{year}{2016}\natexlab{}.
\newblock \showarticletitle{Neural Enquirer: Learning to Query Tables in
  Natural Language}. In \bibinfo{booktitle}{\emph{Proc. of IJCAI '16}}.
  \bibinfo{pages}{2308--2314}.
\newblock


\bibitem[\protect\citeauthoryear{Zhang and Chakrabarti}{Zhang and
  Chakrabarti}{2013}]%
        {Zhang:2013:ISM}
\bibfield{author}{\bibinfo{person}{Meihui Zhang} {and} \bibinfo{person}{Kaushik
  Chakrabarti}.} \bibinfo{year}{2013}\natexlab{}.
\newblock \showarticletitle{InfoGather+: Semantic Matching and Annotation of
  Numeric and Time-varying Attributes in Web Tables}. In
  \bibinfo{booktitle}{\emph{Proc. of SIGMOD '13}}. \bibinfo{pages}{145--156}.
\newblock


\bibitem[\protect\citeauthoryear{Zhang and Balog}{Zhang and Balog}{2017a}]%
        {Zhang:2017:DPF}
\bibfield{author}{\bibinfo{person}{Shuo Zhang} {and} \bibinfo{person}{Krisztian
  Balog}.} \bibinfo{year}{2017}\natexlab{a}.
\newblock \showarticletitle{Design Patterns for Fusion-Based Object Retrieval}.
  In \bibinfo{booktitle}{\emph{Proceedings of the 39th European conference on
  Advances in Information Retrieval}} \emph{(\bibinfo{series}{ECIR '17})}.
  \bibinfo{publisher}{Springer}, \bibinfo{pages}{684--690}.
\newblock


\bibitem[\protect\citeauthoryear{Zhang and Balog}{Zhang and Balog}{2017b}]%
        {Zhang:2017:ESA}
\bibfield{author}{\bibinfo{person}{Shuo Zhang} {and} \bibinfo{person}{Krisztian
  Balog}.} \bibinfo{year}{2017}\natexlab{b}.
\newblock \showarticletitle{EntiTables: Smart Assistance for Entity-Focused
  Tables}. In \bibinfo{booktitle}{\emph{Proc. of SIGIR '17}}.
  \bibinfo{pages}{255--264}.
\newblock


\bibitem[\protect\citeauthoryear{Zhang and Balog}{Zhang and Balog}{2018a}]%
        {Zhang:2018:AHT}
\bibfield{author}{\bibinfo{person}{Shuo Zhang} {and} \bibinfo{person}{Krisztian
  Balog}.} \bibinfo{year}{2018}\natexlab{a}.
\newblock \showarticletitle{Ad Hoc Table Retrieval Using Semantic Similarity}.
  In \bibinfo{booktitle}{\emph{Proceedings of The Web Conference}}
  \emph{(\bibinfo{series}{WWW '18})}. \bibinfo{pages}{1553--1562}.
\newblock


\bibitem[\protect\citeauthoryear{Zhang and Balog}{Zhang and Balog}{2018b}]%
        {Zhang:2018:OTG}
\bibfield{author}{\bibinfo{person}{Shuo Zhang} {and} \bibinfo{person}{Krisztian
  Balog}.} \bibinfo{year}{2018}\natexlab{b}.
\newblock \showarticletitle{On-the-fly Table Generation}. In
  \bibinfo{booktitle}{\emph{Proceedings of 41st International ACM SIGIR
  Conference on Research and Development in Information Retrieval}}.
\newblock


\bibitem[\protect\citeauthoryear{Zwicklbauer, Einsiedler, Granitzer, and
  Seifert}{Zwicklbauer et~al\mbox{.}}{2013}]%
        {Zwicklbauer:2013:TDW}
\bibfield{author}{\bibinfo{person}{Stefan Zwicklbauer},
  \bibinfo{person}{Christoph Einsiedler}, \bibinfo{person}{Michael Granitzer},
  {and} \bibinfo{person}{Christin Seifert}.} \bibinfo{year}{2013}\natexlab{}.
\newblock \showarticletitle{Towards Disambiguating Web Tables}. In
  \bibinfo{booktitle}{\emph{Proc. of ISWC-PD' 13}}. \bibinfo{pages}{205--208}.
\newblock


\end{thebibliography}

\end{document}